\documentclass[12pt]{article}
\usepackage{amsmath}
\usepackage{graphicx,psfrag,epsf}
\usepackage{enumerate}
\usepackage{natbib}
\usepackage{url} 
\usepackage{amsmath,amssymb,amsfonts,amsthm,mathtools} 
\usepackage{bm,bbm}
\usepackage[hidelinks]{hyperref}
\usepackage{color}
\usepackage{subfigure} 
\usepackage{algorithm,algpseudocode}
\usepackage{scalefnt}
\usepackage{rotating}
\usepackage{adjustbox}

\begin{document}

\def\spacingset#1{\renewcommand{\baselinestretch}%
{#1}\small\normalsize} \spacingset{1}


\title{\bf {Bivariate autoregressive conditional models: A new method for jointly modeling duration and number of transactions of irregularly spaced financial data}}

\author{
	\large
	{{Helton Saulo}$^{1,2}$, {Suvra Pal}$^{1}$ and {Roberto Vila}$^{2}$}\\[-0.05cm]
	{\small $^{1}$Department of Mathematics, University of Texas at Arlington, Arlington, TX, USA}\\[-0.05cm]
	{\small $^{2}$Department of Statistics, Universidade de Bras\'{i}lia, Bras\'{i}lia, Brazil}\\[-0.05cm]
}

\date{}

\maketitle

\bigskip
\vspace{-0.5cm}
\begin{abstract}
In this paper, a new approach to bivariate modeling of autoregressive conditional duration (ACD) models is proposed. Specifically, we consider the joint modeling of durations and the number of transactions made during the spell. The proposed bivariate ACD model is based on log-symmetric distributions, which are useful for modeling strictly positive, asymmetric and light- and heavy-tailed data, such as transaction-level high-frequency financial data. A Monte Carlo simulation is performed for the assessment of the estimation method and the evaluation of a form of residuals. A real financial transactions data set is  analyzed in order to illustrate the proposed method.
\end{abstract}

\noindent%

\spacingset{1.5} 

\section{Introduction}\label{sec:01}

In financial markets, the duration between transactions (trades, price changes, etc.) is of great importance as it can translate into a proxy for the existence of information in the market. On the one hand, when there is new information, traders tend to trade more and durations tend to be shorter. Similarly, when there is scarcity of information, durations tend to be longer. In summary, high trade frequency suggests informed trading; see \cite{easeyohara:92}.

Autoregressive conditional duration (ACD) models, introduced by \cite{er:98}, are the standard tools to deal with duration data. This is because duration data is collected irregularly and, therefore, traditional time series techniques are not appropriate here. Besides the irregular nature, duration data have intraday seasonality (activity is lower in the the middle of the trading day than at the start and at the close of the trading day), positive skewness, and an unimodal hazard rate function; see \cite{saulolla:19,sauloetal:23}.

Many generalizations of the original ACD model \citep{er:98} have been proposed in the literature that takes into account aspects such as distribution (e.g., \cite{lslm:14}, \cite{b:10}, \cite{ZHENG201651}, \cite{saulo2017log}), conditional dynamics (e.g., \cite{bg:00}, \cite{fernandesgrammig:06}), and time series properties (e.g., \cite{dz:06} and \cite{chiang:07}). In this regard, interested readers may refer to the literature review by \cite{bhogalvariyam:19}.

This paper explores the joint modeling of durations and the number of transactions made during the spell. Intuitively, shorter the duration, smaller is the number of transactions made during the spell, and vice versa. However, a period with many transactions (regardless of the duration spell) also signals the possible existence of information in the market, that is, the number of transactions, as well as the duration, is also a proxy for the existence of information in the market. We develop new bivariate ACD models based on bivariate log-symmetric distributions, which are useful for modeling strictly positive, asymmetric and light- and heavy-tailed data, such as the transaction-level high-frequency financial data. The results demonstrate that the proposed models are practical and useful for joint modeling of duration and number of transactions.

Bivariate extensions of the original ACD model have been proposed by \cite{engleasger:03}, \cite{mosconiolivetti:05}, and \cite{tan:18}. \cite{engleasger:03} proposed a censored bivariate ACD model that can jointly analyze trade and quote arrivals. In particular, they consider a bivariate vector comprising of two types of durations, the forward recurrence duration and the forward quote duration. The idea is to assess how fast information from trades is taken into account in the formulation of new price quotes. \cite{mosconiolivetti:05} proposed bivariate exponential and Weibull ACD models along the same line as \cite{engleasger:03}, but their models circumvent the drawback of information loss when multiple quote revisions occur without intervening trades. The approach developed by \cite{mosconiolivetti:05} considers a bivariate competing risks framework for both transactions and quotes. Finally, the work of \cite{tan:18} considers bivariate ACD models based on bivariate Birnbaum-Saunders and bivariate Birnbaum-Saunders-Student-$t$ distributions. The approach of \cite{tan:18} takes into account matched durations of two stocks. The proposed methodology differs from those mentioned above, as it considers the joint modeling of durations and the number of transactions made during the spell, which has not yet been considered in the literature. The immediate advantage lies in the possibility of retrieving information that is not usually used (the number of transactions is not generally considered), in addition to serving as an alternative proxy for the existence of information in the market, as durations are usually considered; see \cite{easeyohara:92}.

%
%
%

The rest of this paper is organized as follows. In Section \ref{sec:02}, we introduce the bivariate autoregressive conditional models, which are based on the bivariate log-symmetric distributions. In this section, we also describe some aspects of maximum likelihood estimation of the model parameters and residual analysis. In Section~\ref{sec:03}, we perform a simulation study to evaluate the model parameter recovery and to assess the empirical distribution of residuals. In Section \ref{sec:04}, we use a real financial bivariate data set to illustrate the proposed models and methods. Finally, in Section \ref{sec:05}, we make some concluding remarks.

\section{Bivariate autoregressive conditional models}\label{sec:02}

Let $X_1,\ldots,X_T$ be a sequence of successive recorded times at which market events (trade, price change, volume change, bid-ask range change, etc.) occur. Further, let
\begin{eqnarray}\label{eq:espec}
 Y_{1t}&=&X_t-X_{t-1},\nonumber \\
 Y_{2t}&=&\#\,\text{market events during the spell $[X_{t-1},X_t]$},
\end{eqnarray}
for $t=1,\ldots,T$. Thus, we obtain a sequence of pairs $(Y_{1t},Y_{2t}),\ldots, (Y_{1T},Y_{2T})$, where $Y_{1t}$ is the time elapsed between two successive market events and $Y_{2t}$ is the number of market events in the interval $[X_{t-1},X_t]$.



We assume the joint probability density function (PDF) of $(Y_{1t},Y_{2t})$ to follow a bivariate log-symmetric distribution \citep{vilaetal:22} for $t=1,\ldots, T$. That is, we have
\begin{eqnarray}\label{eq:condpdf}
	f_{Y_1,Y_2}(y_{1t},y_{2t};\boldsymbol{\theta})
	=
	{1\over y_{1t} y_{2t}\sigma_1\sigma_2\sqrt{1-\rho^2}{Z}_{g_c}}\,
	g_c\Biggl(
	{\widetilde{y}_{1t}^2-2\rho\widetilde{y}_{1t}\widetilde{y}_{2t}+\widetilde{y}_{2t}^2
	\over
	1-\rho^2}
	\Biggr),
	\quad
	y_{1t},y_{2t}>0,
\end{eqnarray}
where
$
	\widetilde{y}_{it}
	=
	\log({y_{it}/\eta_{it}})^{1/\sigma_i}, \ i=1,2, \nonumber
$
$\eta_i>0$ are conditional medians, $\sigma_i>0$ are dispersion parameters,  $\rho\in(-1,1)$ is a correlation parameter, $g_c$ is a density generator, and ${Z}_{g_c}>0$ is a normalization constant given by
\begin{align}\label{eq:partitionfunction}
Z_{g_c}
&=
\int_{0}^{\infty}\int_{0}^{\infty}
	{1\over y_{1t}y_{2t}\sigma_1\sigma_2\sqrt{1-\rho^2}}\,
	g_c\Biggl(
{\widetilde{y}_{1t}^2-2\rho\widetilde{y}_{1t}\widetilde{y}_{2t}+\widetilde{y}_{2t}^2
	\over
	1-\rho^2}
\Biggr)\, {\rm d}y_{1t}{\rm d}y_{2t}, \quad t=1,\ldots, T.
\end{align}
We use the notation $(Y_{1t},Y_{2t})^{\top}\sim {\rm BLS}(\sigma_1,\sigma_2,\rho,\sigma_{1t},\sigma_{12t},g_c)$. Note that the density generator $g_c$ leads to different bivariate log-symmetric distributions. Table \ref{table:densgene} presents density generator $g_c$ and the corresponding normalization constant $Z_{g_c}$ for some bivariate log-symmetric distributions. In this table, $\Gamma(t)=\int_0^\infty x^{t-1} \exp(-x) \,{\rm d}x$, $t>0$, is the complete gamma function,
$K_0(u)=\int_0^\infty t^{-1} \exp(-t-{u^2\over 4t}) \,{\rm d}t/2$, $u>0$, is the Bessel function of the third kind, and $\gamma(s,x)=\int_{0}^{x}t^{s-1}\exp(-t)\,{\rm d}t$ is the lower incomplete gamma function. From Table \ref{table:densgene}, we observe that some distributions may involve an extra parameter.

The dynamics of the conditional medians are governed by (for $t=1,\ldots, T$)
\begin{eqnarray}\label{eq:dyn-acd}
\begin{array}{lll}
 \log(\eta_{1t})&=&\displaystyle \omega_{1} + \sum_{j=1}^{p_1}\alpha_{1j} \log(\eta_{1,t-j}) + \sum_{j=1}^{q_1}\beta_{1j}\frac{y_{1,t-j}}{\eta_{1,t-j}}, 
 \\[0,6cm]
\log(\eta_{2t})&=&\displaystyle \omega_{2} + \sum_{j=1}^{p_2}\alpha_{2j} \log(\eta_{2,t-j}) + \sum_{j=1}^{q_2}\beta_{2j}\frac{y_{2,t-j}}{\eta_{2,t-j}},
\end{array}
 \end{eqnarray}
implying the corresponding notation \text{BLS-ACD}($p_1, q_1, p_2, q_2,\bm{\theta}, g_c$), where
$$\bm{\theta} = (\sigma_1,\sigma_2,\rho,\omega_{1},\alpha_{11},\ldots,\alpha_{1p_1},\beta_{11},\ldots,\beta_{1q_1},
\omega_{2},\alpha_{21},\ldots,\alpha_{2p_2},\beta_{21},\ldots,\beta_{2q_2})^{\top}.$$

\begin{table}[!ht]
\small
\label{table:densgene}
	\caption{Density generators $g_c$ and normalization constant $Z_{g_c}$ for some bivariate log-symmetric distributions.}
	\vspace*{0.15cm}
	\centering
	\begin{tabular}{llll}
		\hline
		Bivariate distribution
		& $Z_{g_c}$ & $g_c$ & Extra parameter
		\\ [0.5ex]
		\noalign{\hrule height 1.7pt}
		Log-normal
		& $2\pi$ & $\exp(-x/2)$ & $-$
		\\ [1ex]
		Log-Student-$t$
		& ${{\Gamma({\nu/ 2})}\nu\pi\over{\Gamma({(\nu+2)/ 2})}}$
		& $(1+{x\over\nu})^{-(\nu+2)/ 2}$  &  $\nu>0$
		\\ [1ex]
		Log-hyperbolic
		& ${2\pi (\nu+1)\exp(-\nu)\over \nu^2}$ & $\exp(-\nu\sqrt{1+x})$ &  $\nu>0$
		\\ [1ex]
		Log-Laplace
		& $\pi$  & $K_0(\sqrt{2x})$ & $-$
		\\ [1ex]
		Log-slash
		& ${\pi\over \nu-1}\, 2^{3-\nu\over 2}$ & $ x^{-{\nu+1\over 2}} \gamma({\nu+1\over 2},{x\over 2})$ & $\nu>1$
		\\ [1ex]
		Log-power-exponential
		& ${2^{\nu+1}(1+\nu)\Gamma(1+\nu)}\pi$  & ${\exp\bigl(-{1\over 2}\, x^{1/(1+\nu)}\bigr)}$ & $-1<\nu\leqslant 1$
		\\ [1ex]
		Log-logistic
		& $\pi/2$  & {${\exp(-x)\over (1+\exp(-x))^2}$} &  {$-$}
 	\\ [1ex]
	\hline
	\end{tabular}
\end{table}

\subsection{Maximum likelihood estimation}

Let $\{(Y_{1t},Y_{2t}): t=1,\ldots,T\}$ be a bivariate random sample of size $T$ from the BLS-ACD($p_1, q_1, p_2, q_2; \bm{\theta}, g_c$) model with conditional joint PDF as given in \eqref{eq:condpdf}, and let $(y_{1t},y_{2t})$ be a realization of $(Y_{1t},Y_{2t})$. To obtain the maximum likelihood estimates (MLEs) of the model parameters with parameter vector $\bm{\theta} = (\sigma_1,\sigma_2,\rho,\omega_{1},\alpha_{11},\ldots,\alpha_{1p_1},\beta_{11},\ldots,\beta_{1q_1},
\omega_{2},\alpha_{21},\ldots,\alpha_{2p_2},$ $\beta_{21},\ldots,\beta_{2q_2})^{\top}$, we maximize the following log-likelihood function (without the additive constant):
\begin{align*}
\ell(\boldsymbol{\theta})
=
-T\log(\sigma_1)-T\log(\sigma_2)-{T\over 2}\,\log\big({1-\rho^2}\big)
+
\sum_{t=1}^{T}
\log
g_c(q(\boldsymbol{y}_t)),
\quad 
\boldsymbol{y}_t=({y}_{1t}, {y}_{2t})^\top\in\mathbb{R}^2_+,
\end{align*}
where
$
q(\boldsymbol{y}_t)
=	{(\widetilde{y}_{1t}^2-2\rho\widetilde{y}_{1t}\widetilde{y}_{2t}+\widetilde{y}_{2t}^2)
/
(1-\rho^2)},
$
for $
\widetilde{y}_{it}
=
\log({y_{it}/\eta_{it}})^{1/\sigma_i}, \ i=1,2
$.
{
The likelihood equations are given by (for $i=1,2$)
\begin{align*}
&{\partial\ell(\boldsymbol{\theta})\over\partial\sigma_i}
=
-{T\over \sigma_i}
+
\sum_{t=1}^{T} 
{
g'_c(q(\boldsymbol{y}_t))
\over 
g_c(q(\boldsymbol{y}_t))}
\, 
{\partial q(\boldsymbol{y}_t)\over\partial\sigma_i}=0,
\\[0,2cm]
&{\partial\ell(\boldsymbol{\theta})\over\partial\rho}
=
{\rho T\over 1-\rho^2}
+
\sum_{t=1}^{T} 
{
	g'_c(q(\boldsymbol{y}_t))
	\over 
	g_c(q(\boldsymbol{y}_t))}
\, 
{\partial q(\boldsymbol{y}_t)\over\partial\rho}=0,
\\[0,2cm]
&{\partial\ell(\boldsymbol{\theta})\over\partial\gamma}
=
\sum_{t=1}^{T} 
{
	g'_c(q(\boldsymbol{y}_t))
	\over 
	g_c(q(\boldsymbol{y}_t))}
\, 
{\partial q(\boldsymbol{y}_t)\over\partial\gamma}=0,
\quad
\gamma\in\{\omega_i,\alpha_{ik},\beta_{ik}\}.
\end{align*}
Since
\begin{align}\label{exp-q}
q(\boldsymbol{y}_t)
=
{\big[{\log(y_{1t})-\log(\eta_{1t})\over \sigma_1}\big]^2-2\rho\big[{\log(y_{1t})-\log(\eta_{1t})\over \sigma_1}\big]\big[{\log(y_{2t})-\log(\eta_{2t})\over \sigma_2}\big]+\big[{\log(y_{2t})-\log(\eta_{2t})\over \sigma_2}\big]^2
	\over 
	1-\rho^2},
\end{align}
we have (for $i\neq r=1,2$)
\begin{align}
	&{\partial q(\boldsymbol{y}_t)\over\partial\sigma_i}
	=
	-{2\over\sigma_i}\,
	{ \big[{\log(y_{it})-\log(\eta_{it})\over \sigma_i}\big]^2
	-
	\rho\big[{\log(y_{it})-\log(\eta_{it})\over \sigma_i}\big]
	\big[{\log(y_{rt})-\log(\eta_{rt})\over \sigma_r}\big]
		\over 
		1-\rho^2}, \label{der-1}
	\\[0,2cm]	
	&{\partial q(\boldsymbol{y}_t)\over\partial\rho}
	=
	-2\rho \, 
	{q(\boldsymbol{y}_t) +\big[{\log(y_{1t})-\log(\eta_{1t})\over \sigma_1}\big]\big[{\log(y_{2t})-\log(\eta_{2t})\over \sigma_2}\big]
		\over 
		1-\rho^2}. \label{der-2}
\end{align}
Moreover, by using \eqref{exp-q} and relations in \eqref{eq:dyn-acd}, we obtain (for $i\neq r=1,2$)
\begin{align}
	&{\partial q(\boldsymbol{y}_t)\over\partial\omega_i}
	=
	-{2\over\sigma_i}\,
	{
		\big[{\log(y_{it})-\log(\eta_{it})\over \sigma_i}\big]
		-
		\big[{\log(y_{rt})-\log(\eta_{rt})\over \sigma_r}\big]
		\over 
		1-\rho^2
	}, \label{der-3}
	\\[0,2cm]
	&{\partial q(\boldsymbol{y}_t)\over\partial\alpha_{ik}}
	=
	-{2\over\sigma_i}\,
	{
		\big[{\log(y_{it})-\log(\eta_{it})\over \sigma_i}\big]
		-
		\big[{\log(y_{rt})-\log(\eta_{rt})\over \sigma_r}\big]
		\over 
		1-\rho^2
	}\,
	\log(\eta_{i,t-k}), \label{der-4}
	\\[0,2cm]
	&{\partial q(\boldsymbol{y}_t)\over\partial\beta_{ik}}
	=
	-{2\over\sigma_i}\,
	{
		\big[{\log(y_{it})-\log(\eta_{it})\over \sigma_i}\big]
		-
		\big[{\log(y_{rt})-\log(\eta_{rt})\over \sigma_r}\big]
		\over 
		1-\rho^2
	}\,
	\frac{y_{i,t-k}}{\eta_{i,t-k}}. \label{der-5}
\end{align}
We note that no closed-form solution to the maximization problem is available. As such, the MLE of $\bm\theta$ can only be obtained via numerical optimization. 

As the likelihood function $L=L(\boldsymbol{\theta})$ is
twice continuously differentiable on the parameter space $\Theta$
(assumed to be an open connected subset of $\mathbb{R}^{5+p_1+p_2+q_1+q_2}$),
under the regularity conditions that (i)
the matrix of second partials of $L$ is negative definite for every $\boldsymbol{\theta}\in\Theta$ where the gradient of $L$ vanishes and
(ii) $L$ approaches a constant on the boundary of $\Theta$,
%
\cite{Makelainen1981} established 
the existence and uniqueness of
a maximum likelihood (ML) estimator $\widehat{\boldsymbol{\theta}}$ for $\boldsymbol{\theta}$. Due to the generality of the distributions considered and the large number of parameters in the model, it is a challenge to verify these conditions, which are generally assumed to be valid, since, if they are not valid, it is commonplace to restrict the parameter space $\Theta$ to a compact set where, by extreme value theorem, the continuous function $L$ always will reach a maximum value, guaranteeing the existence of an ML  estimator.
The extra parameter is estimated by using a two-step profile log-likelihood approach; see \cite{sauloetal:22}. 

%
With regard to the asymptotic behavior of the ML estimator $\widehat{\boldsymbol{\theta}}$ of $\boldsymbol{\theta}$,
under some regularity conditions,  $\sqrt{n}(\widehat{\boldsymbol{\theta}}-\boldsymbol{\theta})
\stackrel{d}{\longrightarrow} N(\boldsymbol{0},I^{-1}(\boldsymbol{\theta}))$ \cite[see][]{Davison2008},
where  
$\boldsymbol{0}$ is the zero mean vector and $I^{-1}(\boldsymbol{\theta})$ is the inverse of the expected 
Fisher information matrix.
As is well-known, the aforementioned asymptotic convergence result is useful
for constructing confidence regions and for performing
hypothesis testing related to $\boldsymbol{\theta}$  \cite[see][]{Davison2008}. Due to asymptotic normality of the ML estimates, it is preferable to use the observed Fisher information (the negative of the Hessian matrix) instead of the expected Fisher information \cite[see][]{EfronHinkley:1978}.
By denoting 
$
h(q(\boldsymbol{y}_t))
	=
{
	g'_c(q(\boldsymbol{y}_t))
/
	g_c(q(\boldsymbol{y}_t))},
$
the elements of the Hessian matrix are given by
\begin{align*}
	&{\partial^2\ell(\boldsymbol{\theta})\over\partial\sigma_i\partial\xi}
	=
	{T\over \sigma_i^2}\, \delta_{\sigma_i,\xi}
	+
	\sum_{t=1}^{T} 
	\left\{
	h'(q(\boldsymbol{y}_t))\,
	{\partial q(\boldsymbol{y}_t)\over\partial\sigma_i}\,
	{\partial q(\boldsymbol{y}_t)\over\partial\xi}
	+
	h(q(\boldsymbol{y}_t))
	{\partial^2 q(\boldsymbol{y}_t)\over\partial\sigma_i\partial\xi}
	\right\},
	\\[0,2cm]
	&
	{\partial^2\ell(\boldsymbol{\theta})\over\partial\rho\partial\zeta}
	=
	{(1+\rho^2) T\over (1-\rho^2)^2}\, \delta_{\rho,\zeta}
	+
	\sum_{t=1}^{T} 
	\left\{
h'(q(\boldsymbol{y}_t))\,
{\partial q(\boldsymbol{y}_t)\over\partial\rho}\,
{\partial q(\boldsymbol{y}_t)\over\partial\zeta}
+
h(q(\boldsymbol{y}_t))
{\partial^2 q(\boldsymbol{y}_t)\over\partial\rho\partial\zeta}
\right\},
\\[0,2cm]
&{\partial^2\ell(\boldsymbol{\theta})\over\partial\gamma\partial\varphi}
=
	\sum_{t=1}^{T} 
\left\{
h'(q(\boldsymbol{y}_t))\,
{\partial q(\boldsymbol{y}_t)\over\partial\gamma}\,
{\partial q(\boldsymbol{y}_t)\over\partial\varphi}
+
h(q(\boldsymbol{y}_t))
{\partial^2 q(\boldsymbol{y}_t)\over\partial\gamma\partial\varphi}
\right\},
\end{align*}
for $\xi\in\{\sigma_i,\sigma_l,\rho,\omega_i,\alpha_{ik},\beta_{ik}\}$, $i\neq l=1,2$;  $\zeta\in\{\rho,\omega_i,\alpha_{ik},\beta_{ik}\}$; and $\gamma,\varphi\in\{\omega_i,\alpha_{ik},\beta_{ik}\}$ ($i=1,2$).
Here, $\delta_{i,j}$ denotes the Kronecker delta, which is one if $i=j$, and is zero otherwise.
The second-order partial derivatives in the above equations, i.e., 
\begin{align*}
{\partial^2 q(\boldsymbol{y}_t)\over\partial\sigma_i\partial\xi},
\quad 
{\partial^2 q(\boldsymbol{y}_t)\over\partial\rho\partial\zeta}
\quad \text{and} \quad 
{\partial^2 q(\boldsymbol{y}_t)\over\partial\gamma\partial\varphi}
\end{align*}
can be easily calculated from the first-order derivatives \eqref{der-1}-\eqref{der-5} with the relations in \eqref{eq:dyn-acd}. For the sake of brevity, we omit these expressions.
}

\subsection{Residual analysis}\label{sec:residual}

Goodness-of-fit and departures from the model assumptions are assessed through the following stochastic relation:
\begin{align}\label{eq:residstochas}
\text{Re}(Y_{1t},Y_{2t})
=
		{\widetilde{Y}_{1t}^2-2\rho\widetilde{Y}_{1t}\widetilde{Y}_{2t}+\widetilde{Y}_{2t}^2
		\over
		1-\rho^2},
\end{align}
where
$\widetilde{Y}_{it}=\log({Y_{it}/ \eta_{it}})^{1/\sigma_{i}}$, for $i=1,2$ and $t=1,\ldots,T$, with
$\sigma_{it}$ and $\eta_{it}$ being as in \eqref{eq:condpdf} and \eqref{eq:dyn-acd}, respectively.  It is straightforward to show that
the cumulative distribution function (CDF) and PDF of $\text{Re}(Y_{1t},Y_{2t})$ are given, respectively, by
\begin{eqnarray}\label{eq:resdeq}
	F_{\text{Re}(Y_{1t},Y_{2t})}(x)
   &=&
	{4\over Z_{g_c}}\,
	\int_{0}^{\sqrt{x}}
	\left[\int_{0}^{\sqrt{x-z_1^2}} g_c(z_1^2+z_2^2) \, {\rm d}z_2\right] {\rm d}z_1, \quad  x>0,\\
f_{\text{Re}(Y_{1t},Y_{2t})}(x)&=&{\pi\over Z_{g_c}}\, g_c(x), \quad x>0,
	\label{eq-2}
\end{eqnarray}
where $g_c$ and $Z_{g_c}$ are given in Table~\ref{table:densgene}. Thus, given the reference distribution, it is possible to construct quantile-quantile (QQ) plots to assess the fit.


\section{Monte Carlo simulation}\label{sec:03}

In this section, we present the results of two Monte Carlo simulation studies for the \text{BLS-ACD}($p_1, q_1, p_2, q_2,\bm{\theta}, g_c$) model. The purpose of the first simulation study is to evaluate the performance of the estimators, whereas the second simulation study evaluates the empirical distribution of the residuals defined in Section~\ref{sec:residual}. For illustrative purposes, we only present the results for the bivariate log-normal model for order of the lags $p_1=1$, $q_1=1$, $p_2=1$, $q_2=1$. This is because a higher order did not improve the model fit considerably. Therefore, the dynamics of the conditional medians are governed by
\begin{eqnarray*}
 \log(\eta_{1t})&=& \omega_{1} + \alpha_{11} \log(\eta_{1,t-1}) + \beta_{11}\frac{y_{1,t-1}}{\eta_{1,t-1}}  \ \ \text{and} \\
\log(\eta_{2t})&=& \omega_{2} + \alpha_{21} \log(\eta_{2,t-1}) + \beta_{21}\frac{y_{2,t-1}}{\eta_{2,t-1}}.
 \end{eqnarray*}

The parameter settings for both simulation studies are as follows: sample size $T \in \{500, 1000, 2000\}$, vector of true parameters $(\sigma_1,\sigma_2,\omega_1,\alpha_{11},\beta_{11},\omega_2,\alpha_{21},\beta_{21}) = (1.00,1.00,0.20,0.70,0.10,0.20,0.70,0.10)$, and $\rho\in\{0.10, 0.25, 0.50, 0.75, 0.90\}$. All simulation results are based on $1,000$ Monte Carlo runs.

\subsection{Estimation}\label{sec:03.1}

Tables \ref{tab:MC:bias}-\ref{tab:MC:kurtosis} present the results of Monte Carlo simulation study to assess the performance of the maximum likelihood estimators. In particular, we report the empirical means, biases, root mean squared errors (RMSEs), coefficients of skewness, and coefficients of kurtosis. In terms of parameter recovery, from Tables \ref{tab:MC:bias}-\ref{tab:MC:rmse}, we observe that both biases and RMSEs approach zero as the sample size ($T$) increases; this is the case because the maximum likelihood estimator is consistent. From Tables \ref{tab:MC:bias}-\ref{tab:MC:rmse}, we also observe that both the bias and RMSE tend to be larger for smaller correlation values. The same pattern is observed for negative correlation values, but the results are not reported here for the sake of brevity. The results of Tables \ref{tab:MC:skewness} and \ref{tab:MC:kurtosis} show that all the estimators seem to be consistent and marginally asymptotically distributed as normal. This is because we observe that when $T$ increases, the values of the empirical coefficients of skewness and kurtosis approach to 0 and 3, respectively.

\begin{table}[ht]
\footnotesize
\centering
 \caption{\small {Empirical biases from simulated bivariate log-normal-ACD data.}}\label{tab:MC:bias}
\begin{tabular}{llrrrlrrrlrrr}
  \hline
& & \multicolumn{3}{c}{$\widehat\sigma_1$}&& \multicolumn{3}{c}{$\widehat\sigma_2$} && \multicolumn{3}{c}{$\widehat\rho$}\\
 \cline{3-5} \cline{7-9} \cline{11-13}
$\rho\,\,\,/\,\,\,T$ &  & 500 & 1000 & 2000 &  & 500 & 1000 & 2000 &  & 500 & 1000 & 2000 \\
 \hline
  0.10 &  & -0.0025 & -0.0020 & 0.0000 &  & -0.0049 & -0.0017 & -0.0005 &  & 0.0013 & 0.0011 & 0.0016 \\
  0.25 &  & -0.0039 & -0.0011 & -0.0004 &  & -0.0016 & -0.0016 & -0.0002 &  & 0.0015 & 0.0027 & 0.0006 \\
  0.50 &  & -0.0004 & -0.0026 & -0.0010 &  & -0.0008 & -0.0011 & -0.0012 &  & 0.0033 & 0.0011 & 0.0001 \\
  0.75 &  & -0.0008 & -0.0006 & -0.0012 &  & 0.0000 & -0.0007 & -0.0003 &  & 0.0016 & 0.0011 & -0.0002 \\
  0.90 &  & 0.0004 & -0.0009 & -0.0013 &  & -0.0004 & -0.0012 & -0.0009 &  & 0.0009 & -0.0000 & 0.0001 \\
  \hline
& & \multicolumn{3}{c}{$\widehat\omega_1$}&& \multicolumn{3}{c}{$\widehat\alpha_{11}$} && \multicolumn{3}{c}{$\widehat\beta_{11}$}\\
 \cline{3-5} \cline{7-9} \cline{11-13}
$\rho\,\,\,/\,\,\,T$ &  & 500 & 1000 & 2000 &  & 500 & 1000 & 2000 &  & 500 & 1000 & 2000 \\
 \hline
  0.10 &  & 0.0226 & 0.0105 & 0.0058 &  & -0.0209 & -0.0087 & -0.0058 &  & 0.0006 & 0.0004 & 0.0003 \\
  0.25 &  & 0.0243 & 0.0130 & 0.0055 &  & -0.0221 & -0.0122 & -0.0054 &  & 0.0015 & 0.0006 & 0.0005 \\
  0.50 &  & 0.0272 & 0.0087 & 0.0074 &  & -0.0233 & -0.0077 & -0.0064 &  & 0.0004 & 0.0001 & -0.0001 \\
  0.75 &  & 0.0170 & 0.0038 & 0.0036 &  & -0.0152 & -0.0039 & -0.0031 &  & 0.0004 & 0.0001 & 0.0001 \\
  0.90 &  & 0.0140 & 0.0031 & 0.0037 &  & -0.0115 & -0.0040 & -0.0036 &  & -0.0004 & 0.0005 & 0.0003 \\
  \hline
& & \multicolumn{3}{c}{$\widehat\omega_1$}&& \multicolumn{3}{c}{$\widehat\alpha_{21}$} && \multicolumn{3}{c}{$\widehat\beta_{21}$}\\
 \cline{3-5} \cline{7-9} \cline{11-13}
$\rho\,\,\,/\,\,\,T$ &  & 500 & 1000 & 2000 &  & 500 & 1000 & 2000 &  & 500 & 1000 & 2000 \\
 \hline
  0.10 &  & 0.0248 & 0.0095 & 0.0075 &  & -0.0238 & -0.0083 & -0.0069 &  & 0.0019 & 0.0002 & 0.0001 \\
  0.25 &  & 0.0197 & 0.0072 & 0.0020 &  & -0.0161 & -0.0066 & -0.0024 &  & 0.0001 & 0.0006 & 0.0001 \\
  0.50 &  & 0.0229 & 0.0067 & 0.0051 &  & -0.0199 & -0.0065 & -0.0049 &  & -0.0002 & 0.0005 & 0.0003 \\
  0.75 &  & 0.0152 & 0.0084 & 0.0011 &  & -0.0131 & -0.0084 & -0.0008 &  & -0.0004 & 0.0007 & 0.0002 \\
  0.90 &  & 0.0207 & 0.0055 & 0.0047 &  & -0.0164 & -0.0059 & -0.0043 &  & -0.0008 & 0.0003 & 0.0002 \\
   \hline
\end{tabular}
\end{table}

\begin{table}[ht]
\footnotesize
\centering
 \caption{\small {Empirical RMSEs from simulated bivariate log-normal-ACD data.}}\label{tab:MC:rmse}
\begin{tabular}{llrrrlrrrlrrr}
  \hline
& & \multicolumn{3}{c}{$\widehat\sigma_1$}&& \multicolumn{3}{c}{$\widehat\sigma_2$} && \multicolumn{3}{c}{$\widehat\rho$}\\
 \cline{3-5} \cline{7-9} \cline{11-13}
$\rho\,\,\,/\,\,\,T$ &  & 500 & 1000 & 2000 &  & 500 & 1000 & 2000 &  & 500 & 1000 & 2000 \\
 \hline
  0.10 &  & 0.0314 & 0.0224 & 0.0164 &  & 0.0325 & 0.0228 & 0.0153 &  & 0.0442 & 0.0319 & 0.0225 \\
  0.25 &  & 0.0314 & 0.0227 & 0.0161 &  & 0.0320 & 0.0226 & 0.0166 &  & 0.0437 & 0.0306 & 0.0219 \\
  0.50 &  & 0.0328 & 0.0235 & 0.0155 &  & 0.0313 & 0.0229 & 0.0152 &  & 0.0328 & 0.0244 & 0.0161 \\
  0.75 &  & 0.0325 & 0.0225 & 0.0163 &  & 0.0313 & 0.0229 & 0.0162 &  & 0.0197 & 0.0143 & 0.0096 \\
  0.90 &  & 0.0314 & 0.0223 & 0.0159 &  & 0.0311 & 0.0224 & 0.0160 &  & 0.0084 & 0.0061 & 0.0042 \\
  \hline
& & \multicolumn{3}{c}{$\widehat\omega_1$}&& \multicolumn{3}{c}{$\widehat\alpha_{11}$} && \multicolumn{3}{c}{$\widehat\beta_{11}$}\\
 \cline{3-5} \cline{7-9} \cline{11-13}
$\rho\,\,\,/\,\,\,T$ &  & 500 & 1000 & 2000 &  & 500 & 1000 & 2000 &  & 500 & 1000 & 2000 \\
 \hline
  0.10 &  & 0.1422 & 0.0811 & 0.0525 &  & 0.1248 & 0.0714 & 0.0464 &  & 0.0201 & 0.0139 & 0.0097 \\
  0.25 &  & 0.1255 & 0.0836 & 0.0533 &  & 0.1109 & 0.0750 & 0.0468 &  & 0.0193 & 0.0141 & 0.0095 \\
  0.50 &  & 0.1743 & 0.0691 & 0.0493 &  & 0.1473 & 0.0609 & 0.0430 &  & 0.0186 & 0.0126 & 0.0094 \\
  0.75 &  & 0.0969 & 0.0571 & 0.0392 &  & 0.0849 & 0.0516 & 0.0352 &  & 0.0150 & 0.0110 & 0.0077 \\
  0.90 &  & 0.0892 & 0.0519 & 0.0412 &  & 0.0763 & 0.0460 & 0.0361 &  & 0.0147 & 0.0095 & 0.0066 \\
  \hline
& & \multicolumn{3}{c}{$\widehat\omega_1$}&& \multicolumn{3}{c}{$\widehat\alpha_{21}$} && \multicolumn{3}{c}{$\widehat\beta_{21}$}\\
 \cline{3-5} \cline{7-9} \cline{11-13}
$\rho\,\,\,/\,\,\,T$ &  & 500 & 1000 & 2000 &  & 500 & 1000 & 2000 &  & 500 & 1000 & 2000 \\
 \hline
  0.10 &  & 0.1579 & 0.0769 & 0.0543 &  & 0.1303 & 0.0683 & 0.0483 &  & 0.0208 & 0.0144 & 0.0097 \\
  0.25 &  & 0.1455 & 0.0746 & 0.0522 &  & 0.1232 & 0.0659 & 0.0466 &  & 0.0203 & 0.0135 & 0.0095 \\
  0.50 &  & 0.1572 & 0.0680 & 0.0474 &  & 0.1320 & 0.0600 & 0.0423 &  & 0.0182 & 0.0119 & 0.0086 \\
  0.75 &  & 0.0989 & 0.0583 & 0.0396 &  & 0.0866 & 0.0520 & 0.0346 &  & 0.0149 & 0.0109 & 0.0074 \\
  0.90 &  & 0.1589 & 0.0519 & 0.0438 &  & 0.1275 & 0.0458 & 0.0379 &  & 0.0152 & 0.0096 & 0.0067 \\
   \hline
\end{tabular}
\end{table}

\begin{table}[ht]
\footnotesize
\centering
 \caption{\small {Empirical coefficients of skewness from simulated bivariate log-normal-ACD data.}}\label{tab:MC:skewness}
\begin{tabular}{llrrrlrrrlrrr}
  \hline
& & \multicolumn{3}{c}{$\widehat\sigma_1$}&& \multicolumn{3}{c}{$\widehat\sigma_2$} && \multicolumn{3}{c}{$\widehat\rho$}\\
 \cline{3-5} \cline{7-9} \cline{11-13}
$\rho\,\,\,/\,\,\,T$ &  & 500 & 1000 & 2000 &  & 500 & 1000 & 2000 &  & 500 & 1000 & 2000 \\
 \hline
  0.10 &  & 0.1508 & -0.0023 & 0.0539 &  & -0.0276 & 0.0052 & -0.1072 &  & -0.0142 & 0.0492 & 0.0280 \\
  0.25 &  & 0.0577 & 0.0335 & -0.1006 &  & -0.0039 & 0.0437 & 0.0191 &  & -0.1242 & -0.0348 & -0.1385 \\
  0.50 &  & 0.0177 & -0.0638 & 0.0674 &  & -0.1074 & -0.0097 & 0.0992 &  & -0.0009 & -0.1254 & -0.0504 \\
  0.75 &  & -0.0093 & 0.0092 & 0.1496 &  & 0.1247 & -0.0113 & -0.0408 &  & -0.2656 & -0.1013 & -0.0850 \\
  0.90 &  & 0.0791 & -0.0901 & 0.0608 &  & 0.0769 & 0.0310 & -0.0550 &  & -0.1305 & -0.0921 & -0.2372 \\
  \hline
& & \multicolumn{3}{c}{$\widehat\omega_1$}&& \multicolumn{3}{c}{$\widehat\alpha_{11}$} && \multicolumn{3}{c}{$\widehat\beta_{11}$}\\
 \cline{3-5} \cline{7-9} \cline{11-13}
$\rho\,\,\,/\,\,\,T$ &  & 500 & 1000 & 2000 &  & 500 & 1000 & 2000 &  & 500 & 1000 & 2000 \\
 \hline
  0.10 &  & 3.4462 & 0.4381 & 0.5787 &  & -2.9853 & -0.3418 & -0.4210 &  & -0.0221 & 0.0028 & 0.1451 \\
  0.25 &  & 1.6008 & 1.7952 & 0.6188 &  & -1.6001 & -1.7271 & -0.5204 &  & 0.1339 & 0.0428 & -0.0041 \\
  0.50 &  & 6.6192 & 0.6152 & 0.5166 &  & -5.8546 & -0.5352 & -0.4351 &  & -0.3061 & -0.0741 & -0.8817 \\
  0.75 &  & 1.4306 & 0.7163 & 0.3954 &  & -1.3458 & -0.6481 & -0.3154 &  & -0.0383 & -0.0951 & 0.0903 \\
  0.90 &  & 3.1708 & 0.6222 & 5.2778 &  & -2.2010 & -0.4994 & -5.1252 &  & -0.3232 & -0.0893 & 0.2092 \\
  \hline
& & \multicolumn{3}{c}{$\widehat\omega_1$}&& \multicolumn{3}{c}{$\widehat\alpha_{21}$} && \multicolumn{3}{c}{$\widehat\beta_{21}$}\\
 \cline{3-5} \cline{7-9} \cline{11-13}
$\rho\,\,\,/\,\,\,T$ &  & 500 & 1000 & 2000 &  & 500 & 1000 & 2000 &  & 500 & 1000 & 2000 \\
 \hline
  0.10 &  & 6.1403 & 0.6243 & 0.4031 &  & -4.3236 & -0.4892 & -0.3865 &  & 0.0384 & 0.0796 & 0.1767 \\
  0.25 &  & 6.0516 & 0.5778 & 0.6676 &  & -4.8873 & -0.4590 & -0.6564 &  & -0.0729 & 0.1001 & 0.0692 \\
  0.50 &  & 7.0216 & 0.6751 & 0.4522 &  & -5.7743 & -0.6197 & -0.4211 &  & -0.2838 & 0.0406 & 0.1147 \\
  0.75 &  & 1.1847 & 0.5788 & 0.3379 &  & -1.0465 & -0.5382 & -0.2489 &  & 0.1435 & -0.0514 & 0.0972 \\
  0.90 &  & 10.1918 & 0.4559 & 5.3430 &  & -9.0191 & -0.3878 & -5.2463 &  & -0.6824 & -0.0024 & 0.0819 \\
   \hline
\end{tabular}
\end{table}

\begin{table}[ht]
\footnotesize
\centering
 \caption{\small {Empirical coefficients of kurtosis from simulated bivariate log-normal-ACD data.}}\label{tab:MC:kurtosis}
\begin{tabular}{llrrrlrrrlrrr}
  \hline
& & \multicolumn{3}{c}{$\widehat\sigma_1$}&& \multicolumn{3}{c}{$\widehat\sigma_2$} && \multicolumn{3}{c}{$\widehat\rho$}\\
 \cline{3-5} \cline{7-9} \cline{11-13}
$\rho\,\,\,/\,\,\,T$ &  & 500 & 1000 & 2000 &  & 500 & 1000 & 2000 &  & 500 & 1000 & 2000 \\
 \hline
  0.10 &  & 3.1173 & 3.0355 & 3.0053 &  & 2.9903 & 3.0813 & 3.2384 &  & 2.8016 & 3.0712 & 3.0278 \\
  0.25 &  & 2.8184 & 3.0664 & 3.2155 &  & 3.0669 & 3.3147 & 3.1481 &  & 2.7108 & 2.8990 & 3.0922 \\
  0.50 &  & 2.8979 & 2.8517 & 3.0771 &  & 3.3030 & 2.8923 & 3.0259 &  & 2.9928 & 2.9928 & 2.8801 \\
  0.75 &  & 2.7729 & 3.0627 & 3.0637 &  & 3.0674 & 2.8825 & 2.8547 &  & 2.8790 & 2.7913 & 2.9214 \\
 0.90  &  & 2.9077 & 3.2719 & 2.9461 &  & 2.9328 & 3.1486 & 3.0234 &  & 2.9952 & 2.7802 & 2.7992 \\
  \hline
& & \multicolumn{3}{c}{$\widehat\omega_1$}&& \multicolumn{3}{c}{$\widehat\alpha_{11}$} && \multicolumn{3}{c}{$\widehat\beta_{11}$}\\
 \cline{3-5} \cline{7-9} \cline{11-13}
$\rho\,\,\,/\,\,\,T$ &  & 500 & 1000 & 2000 &  & 500 & 1000 & 2000 &  & 500 & 1000 & 2000 \\
 \hline
  0.10 &  & 32.5862 & 3.4998 & 3.5574 &  & 25.9211 & 3.4601 & 3.3811 &  & 2.9667 & 3.0268 & 2.8191 \\
  0.25 &  & 8.7904 & 14.7759 & 3.9104 &  & 9.5278 & 15.3171 & 3.8924 &  & 3.1507 & 3.2600 & 2.9169 \\
  0.50 &  & 72.8553 & 3.8541 & 3.5183 &  & 61.3777 & 4.3538 & 3.3830 &  & 4.3025 & 3.1762 & 12.4577 \\
  0.75 &  & 9.1342 & 4.1955 & 4.1677 &  & 8.8404 & 4.4009 & 3.9754 &  & 3.3498 & 3.0749 & 3.1441 \\
 0.90  &  & 7.4357 & 3.6893 & 3.1211 &  & 6.0091 & 3.1837 & 3.0914 &  & 2.8949 & 3.3279 & 3.2810 \\
  \hline
& & \multicolumn{3}{c}{$\widehat\omega_1$}&& \multicolumn{3}{c}{$\widehat\alpha_{21}$} && \multicolumn{3}{c}{$\widehat\beta_{21}$}\\
 \cline{3-5} \cline{7-9} \cline{11-13}
$\rho\,\,\,/\,\,\,T$ &  & 500 & 1000 & 2000 &  & 500 & 1000 & 2000 &  & 500 & 1000 & 2000 \\
 \hline
  0.10 &  & 77.1384 & 3.7038 & 3.1719 &  & 45.4248 & 3.6874 & 3.2715 &  & 3.7535 & 2.9378 & 3.3774 \\
  0.25 &  & 81.0042 & 3.3519 & 4.2042 &  & 60.6718 & 3.1812 & 4.1652 &  & 3.6248 & 3.0032 & 3.1402 \\
  0.50 &  & 88.6354 & 3.6941 & 3.8968 &  & 67.6320 & 3.7229 & 3.8107 &  & 4.4525 & 3.2663 & 3.3894 \\
  0.75 &  & 6.3850 & 3.9259 & 3.1587 &  & 6.0465 & 4.4050 & 3.0560 &  & 3.1113 & 3.2040 & 3.0794 \\
 0.90  &  & 9.3927 & 3.8569 & 3.2366 &  & 6.2744 & 3.3990 & 3.0679 &  & 3.3624 & 3.4624 & 3.1349 \\
   \hline
\end{tabular}
\end{table}

\subsection{Empirical distribution of the residuals}\label{sec:03.1}

Here, we present simulation study results that evaluates the performance of the residuals defined in Section \ref{sec:residual}. We only present the bivariate log-normal-ACD case. Note that for this case the stochastic relation \eqref{eq:residstochas} has PDF (by taking $g_c(x)=\exp(-x/2)$ and $Z_{g_c}=2\pi$, see Table \ref{table:densgene}) given by
\begin{align*}
		f_{\text{Re}(Y_{1t},Y_{2t})}(x)
=
{1\over 2}\, \exp\biggl(-{x\over 2}\biggr)
=
{1\over 2^{k/2}\Gamma(k/2)}\, x^{(k/2)-1} \exp\biggl(-{x\over 2}\biggr),
\quad k=2.
	\end{align*}
Thus, $\text{Re}(Y_{1t},Y_{2t})$ has a chi-squared distribution with $2$ degrees of freedom, i.e., $\text{Re}(Y_{1t},Y_{2t})\sim \chi^2_2$. Table \ref{tab:MC:resid} presents the empirical mean, standard deviation, coefficient of skewness and coefficient of kurtosis, whose values are expected to be 2, 2, 2 and 9, respectively, for $\text{Re}(Y_{1t},Y_{2t})$ in the bivariate log-normal case. From Table \ref{tab:MC:resid}, we note that the residuals considered are generally consistent with reference distribution. Therefore, we can use the $\text{Re}(Y_{1t},Y_{2t})$ residuals to check for the suitability of the proposed model.

\begin{table}[ht]
\footnotesize
\centering
 \caption{\small {Summary statistics of the $\text{Re}(Y_{1t},Y_{2t})$ residuals.}}\label{tab:MC:resid}
\begin{tabular}{llrrrlrrrlrrr}
  \hline
& & \multicolumn{3}{c}{$\rho=0.10$}&& \multicolumn{3}{c}{$\rho=0.25$} \\
 \cline{3-5} \cline{7-9}
$\text{Statistic}\,\,\,/\,\,\,n$ &  & 500 & 1000 & 2000 &  & 500 & 1000 & 2000  \\
 \hline
Mean                    &  &  2.0000 & 2.0000 & 2.0000 & & 2.0000 & 2.0000 & 2.0000 \\
Standard deviation      &  &  1.9922 & 1.9970 & 1.9957 & & 1.9926 & 1.9972 & 1.9988 \\
Coefficient of skewness &  &  1.9308 & 1.9692 & 1.9748 & & 1.9322 & 1.9669 & 1.9771 \\
Coefficient of kurtosis &  &  8.3587 & 8.7119 & 8.7663 & & 8.3773 & 8.6797 & 8.7540 \\[.5ex]
  \hline
& & \multicolumn{3}{c}{$\rho=0.50$}&& \multicolumn{3}{c}{$\rho=0.75$} \\
 \cline{3-5} \cline{7-9}
$\text{Statistic}\,\,\,/\,\,\,n$ &  & 500 & 1000 & 2000 &  & 500 & 1000 & 2000  \\
 \hline
Mean                     &  &  2.0000 & 2.0000 & 2.0000 & & 2.0000 & 2.0000 & 2.0000 \\
Standard deviation       &  &  1.9905 & 1.9941 & 1.9990 & & 1.9922 & 1.9964 & 1.9948 \\
Coefficient of skewness  &  &  1.9269 & 1.9551 & 1.9883 & & 1.9316 & 1.9554 & 1.9716 \\
Coefficient of kurtosis  &  &  8.3214 & 8.5716 & 8.9538 & & 8.4013 & 8.5448 & 8.7661 \\
  \hline
& & \multicolumn{3}{c}{$\rho=0.90$} \\
 \cline{3-5}
$\text{Statistic}\,\,\,/\,\,\,n$ &  & 500 & 1000 & 2000  \\
 \hline
 Mean                    &  & 2.0000 & 2.0000 & 2.0000  \\
Standard deviation       &  & 1.9900 & 1.9950 & 1.9984 \\
Coefficient of skewness  &  & 1.9253 & 1.9644 & 1.9757  \\
Coefficient of kurtosis  &  & 8.3430 & 8.6558 & 8.7419
 \\
   \hline
\end{tabular}
\end{table}


\section{{Application to financial data}}\label{sec:04}

In this section, a real high frequency financial data set is analyzed that corresponds to bid-ask-range durations of Bank of America Corporation (BAC) stock on 15th March 2023. Consider $X_1,\ldots,X_T$ as a sequence of successive recorded times at which bid-ask-range change occur. Then, we have
\begin{eqnarray}\label{eq:espec}
 Y_{1t}&=&X_t-X_{t-1},\nonumber \\
 Y_{2t}&=&\#\,\text{market events during the spell $[X_{t-1},X_t]$},
\end{eqnarray}
for $t=1,\ldots,T$. Here, we define the bid-ask range as
\begin{equation}\label{eq:bidask}
 R_t= [\log(\text{Ask}_t)-\log(\text{Bid}_t)]\times 100,
\end{equation}
where $\text{Ask}_t$ denotes the lowest price at which a seller is willing to sell at time $t$ and $\text{Bid}_t$ denotes the highest price at which a buyer is willing to pay at time $t$. Usually, the highest and lowest price of an asset is used in \eqref{eq:bidask}, thus obtaining the the price range, which is an alternative approach for modeling volatility; see \cite{leaoetal:20}. Therefore, $Y_{1t}$ is the bid-ask-range change duration and $Y_{2t}$ is the corresponding number of market events during the change.

Table \ref{tab:descp_bac1} shows descriptive statistics of both raw and adjusted durations/number of market events. For both data sets, data adjustment was applied to eliminate the intraday seasonalities since the activity is low during lunch and high during the beginning and closing of the trading day. The \texttt{R} package \texttt{ACDm} was used to eliminate the seasonality.
From Table \ref{tab:descp_bac1}, we can see that the median of both series is larger than the mean. In addition, both data sets show positive skewness and a high degree of kurtosis.

\begin{table}[!ht]
\footnotesize
\centering
\caption{{Summary statistics for the BAC data.}}\label{tab:descp_bac1}
\begin{tabular}{lcccccccc}
\hline
                                   &\multicolumn{3}{c}{$Y_{1t}$}      && \multicolumn{3}{c}{$Y_{2t}$}\\
                                   \cline{2-4} \cline{6-8}
                                   &    Plain     &  &   Adjusted     &&     Plain     &  &   Adjusted     \\
\hline
$n$                                &   1708       &  &  1708          &&   1708       &  &  1708           \\
Minimum                            &  1           &  &  0.117         &&   1          &  &   0.304\\
10th percentile                    &  1           &  &  0.135         &&   1          &  &  0.389 \\
Mean                               &  7.364       &  &  1.054         &&   2.577      &  &  1.009 \\
50th percentile (median)           &  4           &  &  0.579         &&   2          &  &  0.820 \\
90th percentile                    &  18          &  &  2.572         &&   4          &  &   1.677 \\
Maximum                            &  86          &  & 10.948         &&   18         &  &  6.916 \\
Standard deviation                 & 9.798        &  &  1.336         &&  1.815       &  & 0.689 \\
Coefficient of variation           & 133.049\%    &  &  126.806\%     &&  70.425      &  &  67.992 \\
Coefficient of skewness            & 3.095        &  &  2.834         &&  3.194       &  &  3.044 \\
Coefficient of excess kurtosis     & 13.080       &  & 10.443         &&  15.601      &  &  14.590\\
\hline
\end{tabular}
\end{table}

\begin{figure}[!ht]
\centering
\subfigure[$Y_{1t},Y_{2t}$]{\includegraphics[height=5.2cm,width=5.2cm]{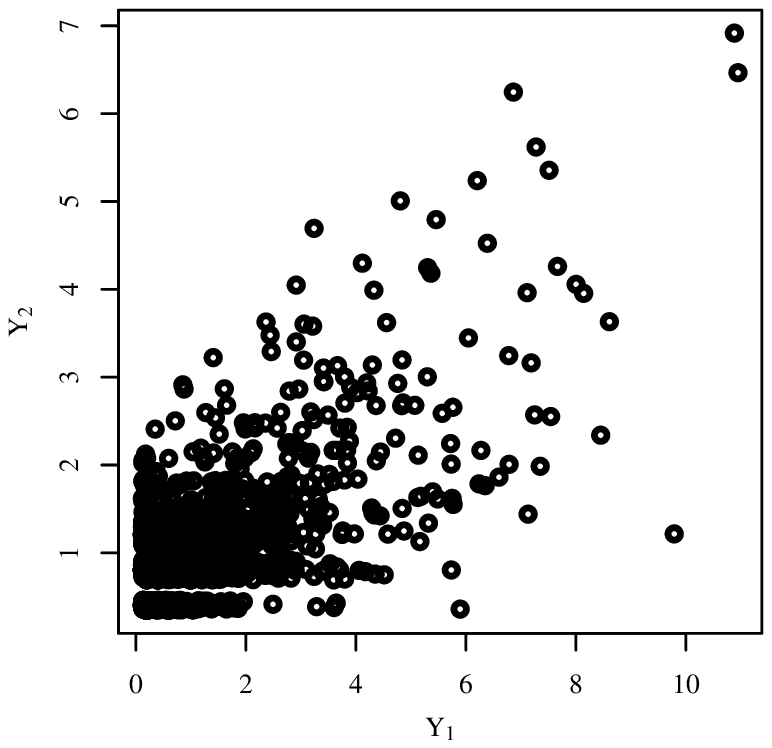}}
\subfigure[$Y_{1t}$]{\includegraphics[height=5.2cm,width=5.2cm]{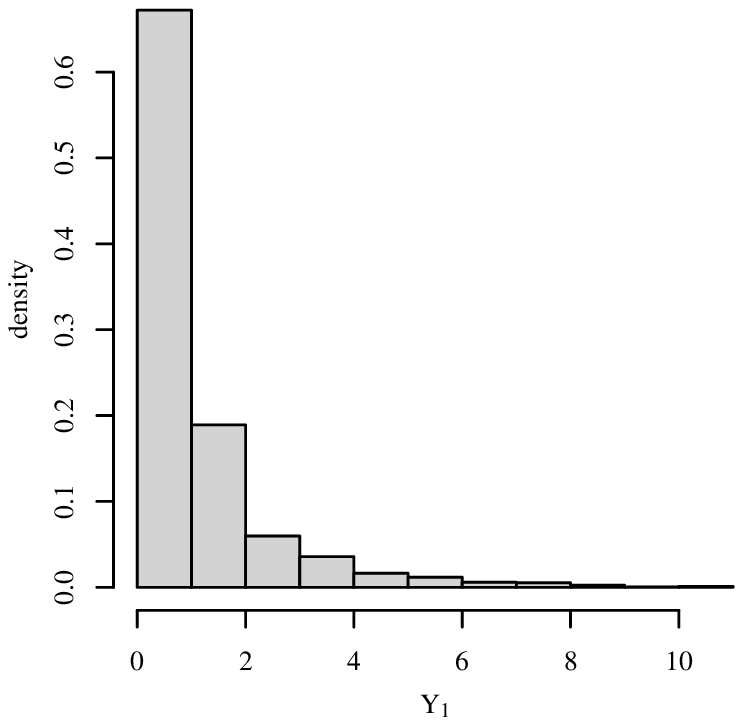}}
\subfigure[$Y_{2t}$]{\includegraphics[height=5.2cm,width=5.2cm]{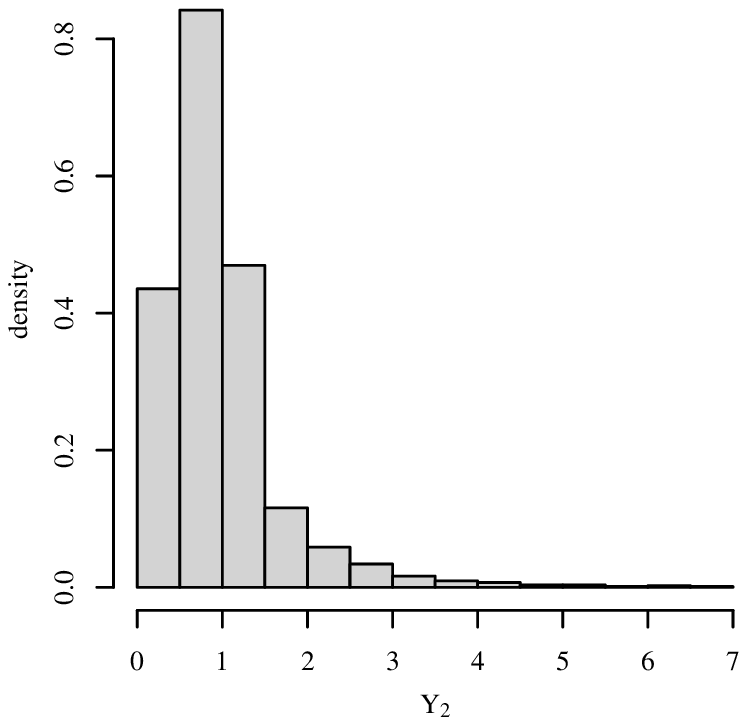}}
 \caption{\small {Scatterplot and histograms for the BAC data.}}
\label{fig:scatterhist}
\end{figure}

Figure \ref{fig:scatterhist} presents the scatterplot and histograms for the seasonally adjusted BAC data. From the scatterplot, we notice a positive correlation, as expected, and the histograms confirm the asymmetric nature of the data. Thus, the proposed bivariate ACD models are good candidates for these data since they account for asymmetricity.

The values of the log-likelihood, and Akaike (AIC) and Bayesian information (BIC) criteria for the adjusted bivariate log-symmetric ACD models are reported in Table \ref{tab:fit_bac1}. Moreover, Table \ref{tab:estima_bac1} presents the maximum likelihood estimates and  the standard errors (SEs) of the model parameters. In general, the bivariate ACD model based on the log-hyperbolic distribution provides the best fit to the data\footnote{The estimated extra parameters for the log-hyperbolic model was $\widehat{\nu}=48$.}.

\begin{table}[ht]
\footnotesize
\centering
 \caption{{Model selection and goodness-of-fit measures for the BAC data.}}\label{tab:fit_bac1}
\begin{tabular}{lllll}
  \hline
Distribution & Log-likelihood & AIC & BIC & CAIC \\
  \hline
  Log-normal & -2503.3361 & 5024.6721 & 5073.6598 & 5024.7781 \\
  Log-Student-$t$ & -2463.8961 & 4945.7922 & 4994.7799 & 4945.8983 \\
  Log-hyperbolic & -2457.6600 & 4933.3200 & 4982.3077 & 4933.4260 \\
  Log-Laplace & -2636.8564 & 5291.7128 & 5340.7005 & 5291.8188 \\
  Log-slash & -2463.8978 & 4945.7956 & 4994.7833 & 4945.9016 \\
  Log-power-exponential & -2469.5302 & 4957.0604 & 5006.0481 & 4957.1664 \\
  Log-logistic & -2472.4021 & 4962.8042 & 5011.7919 & 4962.9102 \\
   \hline
\end{tabular}
\end{table}

\begin{table}[!ht]
\footnotesize
\centering
 \caption{Maximum likelihood estimates with SEs for the BAC data.}\label{tab:estima_bac1}
\begin{tabular}{llrrrrrrrrrrrrrrr}
  \hline
Model & Result & $\widehat\sigma_1$ & $\widehat\sigma_2$ & $\widehat\rho$ & $\widehat\omega_{1}$ & $\widehat\alpha_{11}$ & $\widehat\beta_{11}$ & $\widehat\omega_2$ & $\widehat\alpha_{21}$ & $\widehat\beta_{21}$ \\
  \hline
Log-normal & Estimate & 1.0690 & 0.5480 & 0.5139 & -0.0466 & 0.9522 & 0.0119 & -0.2915 & -0.6606 & 0.0338 \\
   & SE & 0.0183 & 0.0094 & 0.0181 & 0.0195 & 0.0284 & 0.0041 & 0.0267 & 0.1345 & 0.0121 \\[0.7ex]
  Log-Stud. & Estimate & 1.0071 & 0.5265 & 0.4967 & -0.9134 & -0.8641 & -0.0556 & -0.2926 & -0.8951 & -0.0067 \\
   & SE & 0.0179 & 0.0095 & 0.019 & 0.0519 & 0.0267 & 0.0071 & 0.0321 & 0.0960 & 0.0068 \\ [0.7ex]
  Log-hyperb. & Estimate & 7.0347 & 3.7007 & 0.4987 & -0.9016 & -0.8637 & -0.0552 & -0.2840 & -0.8980 & -0.0067 \\
   & SE & 0.1218 & 0.0643 & 0.0184 & 0.0516 & 0.0267 & 0.0071 & 0.0315 & 0.0926 & 0.0067 \\ [0.7ex]
  Log-Laplace & Estimate & 1.2035 & 0.5915 & 0.4679 & -0.9027 & -0.8216 & -0.0640 & -0.3299 & -0.8465 & -0.0111 \\
   & SE & 0.0285 & 0.0143 & 0.0232 & 0.0072 & 0.0063 & 0.0017 & 0.0023 & 0.0063 & 0.0015 \\ [0.7ex]
  Log-slash & Estimate & 0.9081 & 0.475 & 0.4961 & -0.9123 & -0.8564 & -0.0546 & -0.1428 & 0.0529 & -0.0069 \\
   & SE & 0.0161 & 0.0086 & 0.019 & 0.0520 & 0.0274 & 0.0072 & 0.1334 & 0.8215 & 0.0146 \\ [0.7ex]
  Log-power-exp. & Estimate & 0.8216 & 0.4266 & 0.4942 & -0.9106 & -0.8649 & -0.0564 & -0.2935 & -0.8926 & -0.0080 \\
   & SE & 0.0151 & 0.0079 & 0.0194 & 0.0521 & 0.0268 & 0.0072 & 0.0311 & 0.0863 & 0.0069 \\ [0.7ex]
  Log-logistic & Estimate & 1.2400 & 0.6720 & 0.5046 & -0.8684 & -0.8535 & -0.0512 & -0.2583 & -0.9011 & -0.0036 \\
   & SE & 0.0187 & 0.0101 & 0.0163 & 0.0498 & 0.0284 & 0.0068 & 0.034 & 0.1429 & 0.0062 \\
   \hline
\end{tabular}
\end{table}

\begin{figure}[!ht]
\centering
\subfigure[Log-normal]{\includegraphics[height=5cm,width=5cm]{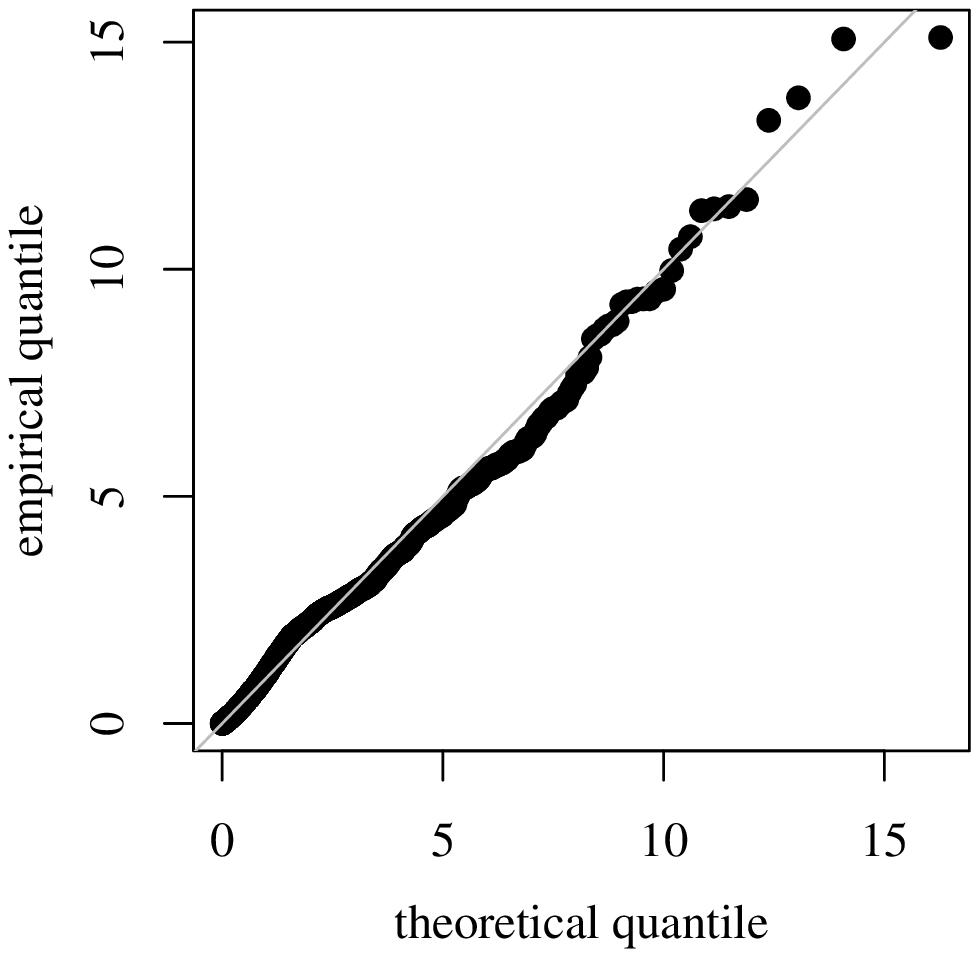}}
\subfigure[Log-Student-$t$]{\includegraphics[height=5cm,width=5cm]{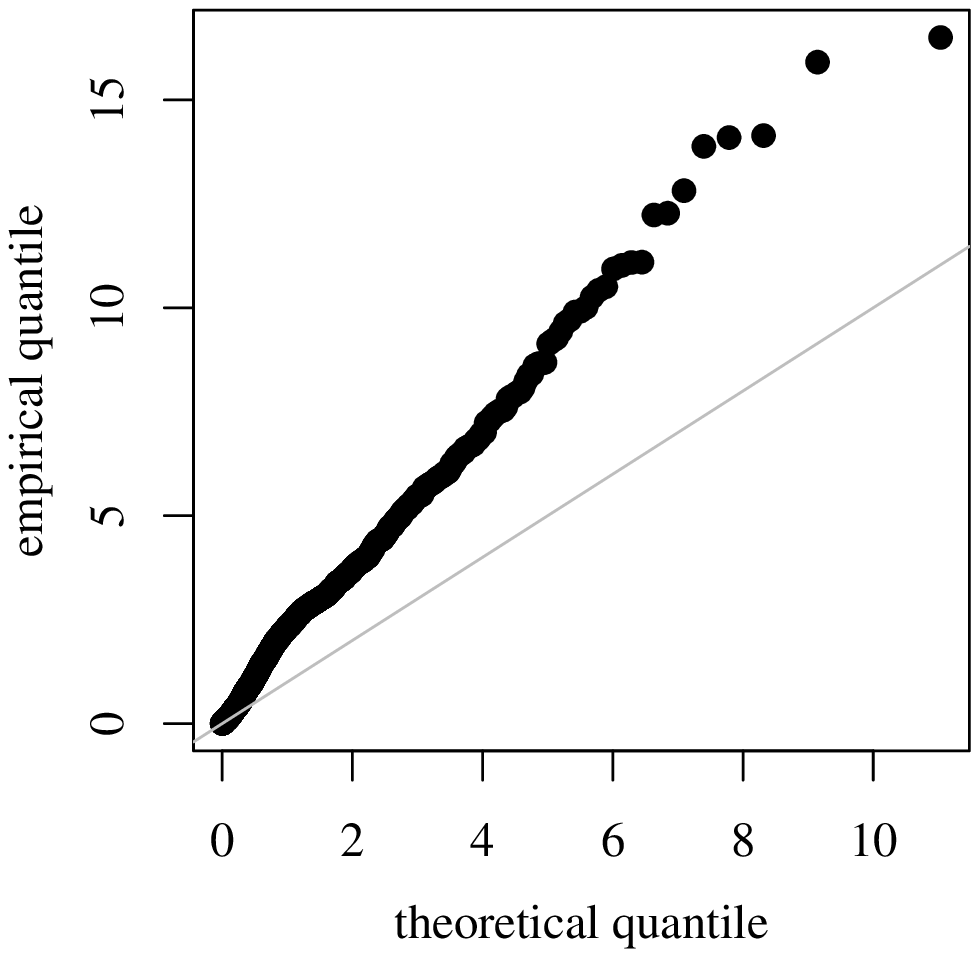}}
\subfigure[Log-hyperbolic]{\includegraphics[height=5cm,width=5cm]{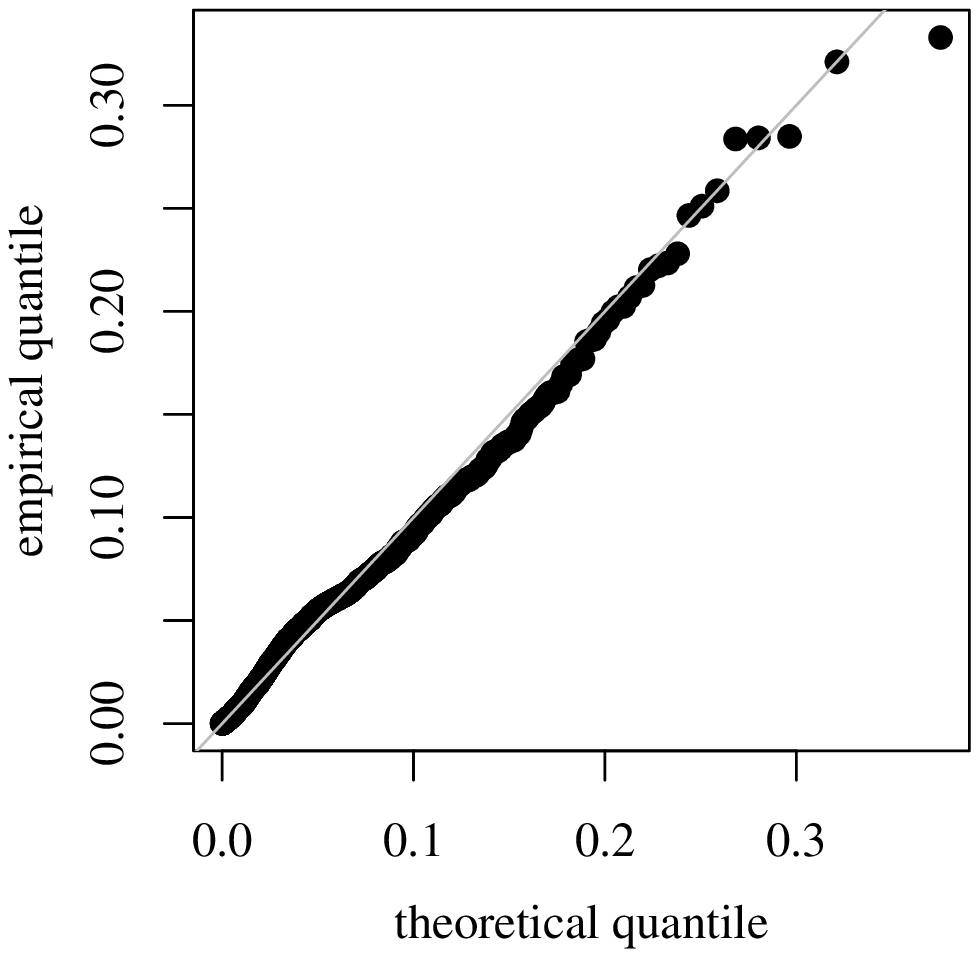}}
\subfigure[Log-Laplace]{\includegraphics[height=5cm,width=5cm]{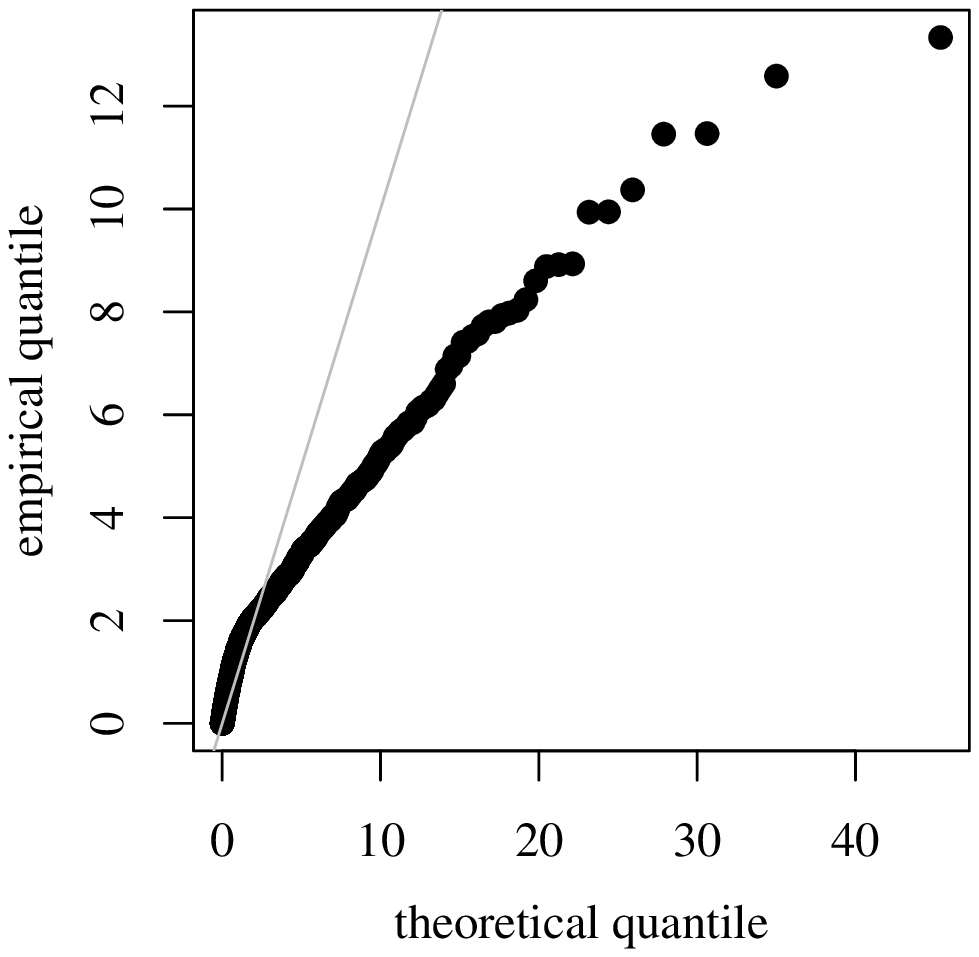}}
\subfigure[Log-slash]{\includegraphics[height=5cm,width=5cm]{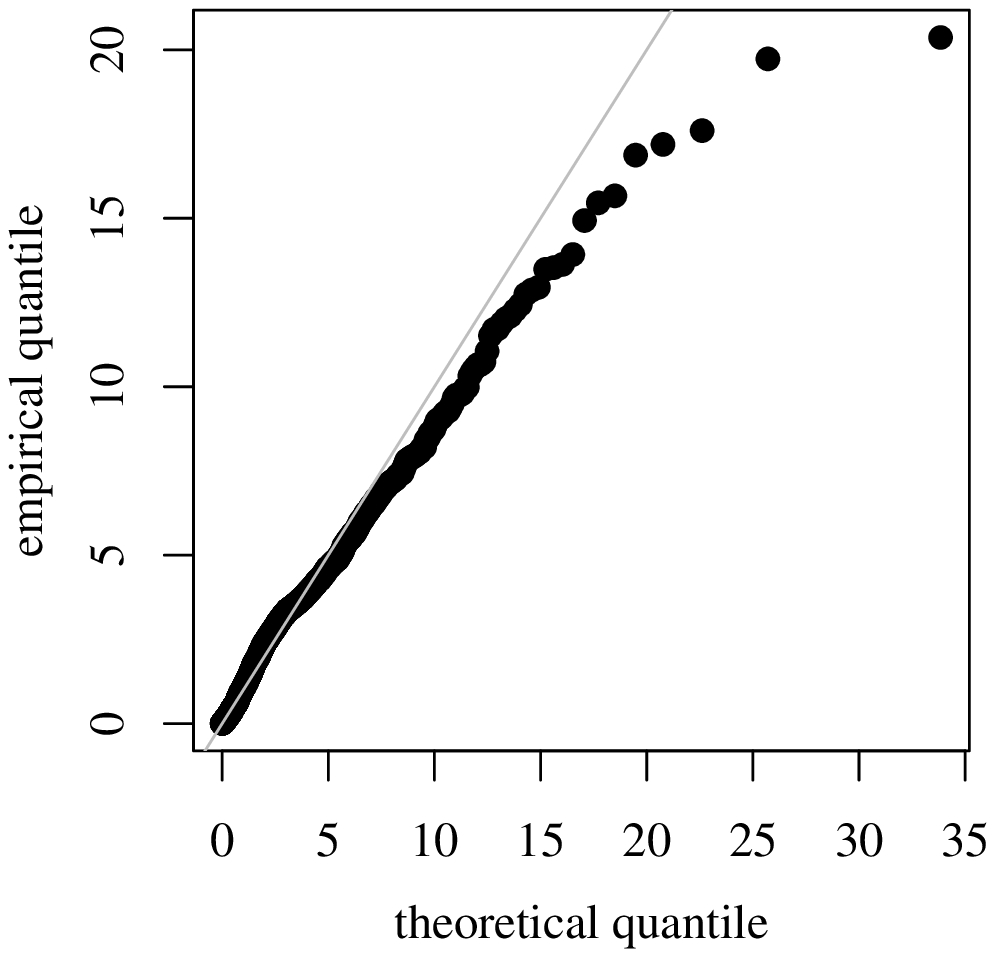}}
\subfigure[Log-power-exponential]{\includegraphics[height=5cm,width=5cm]{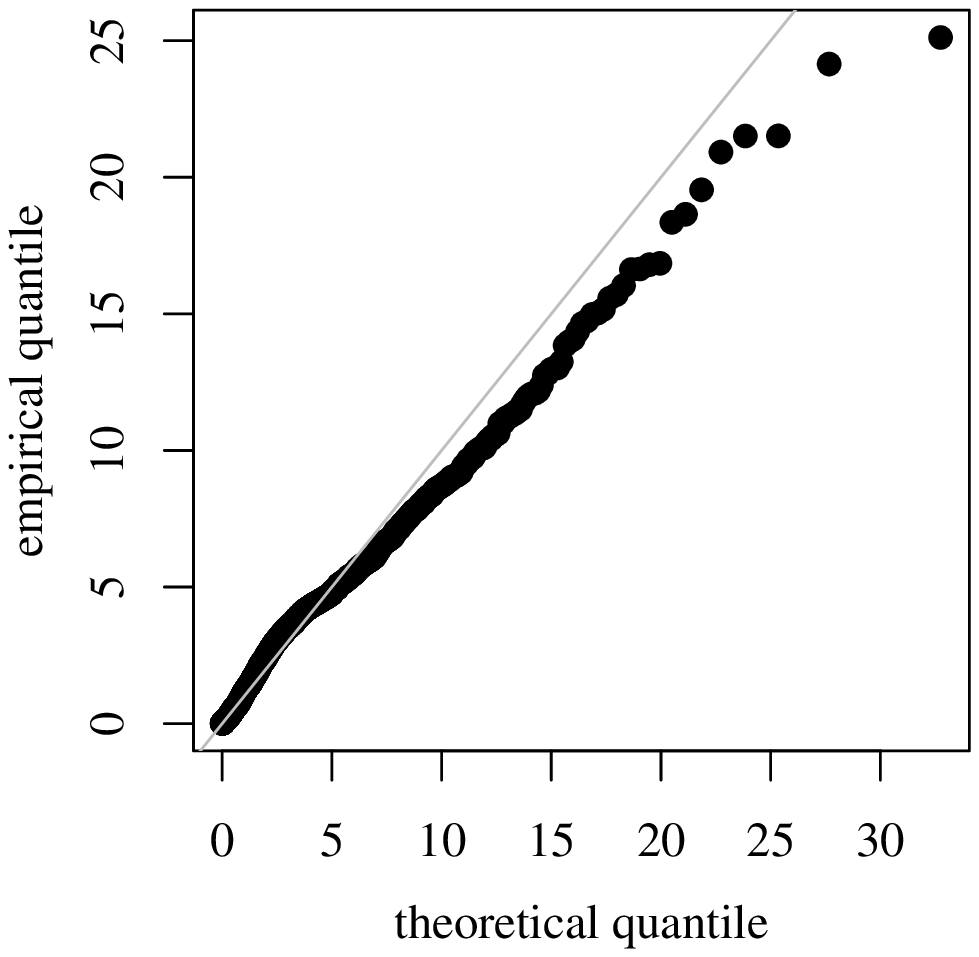}}
\subfigure[Log-logistic]{\includegraphics[height=5cm,width=5cm]{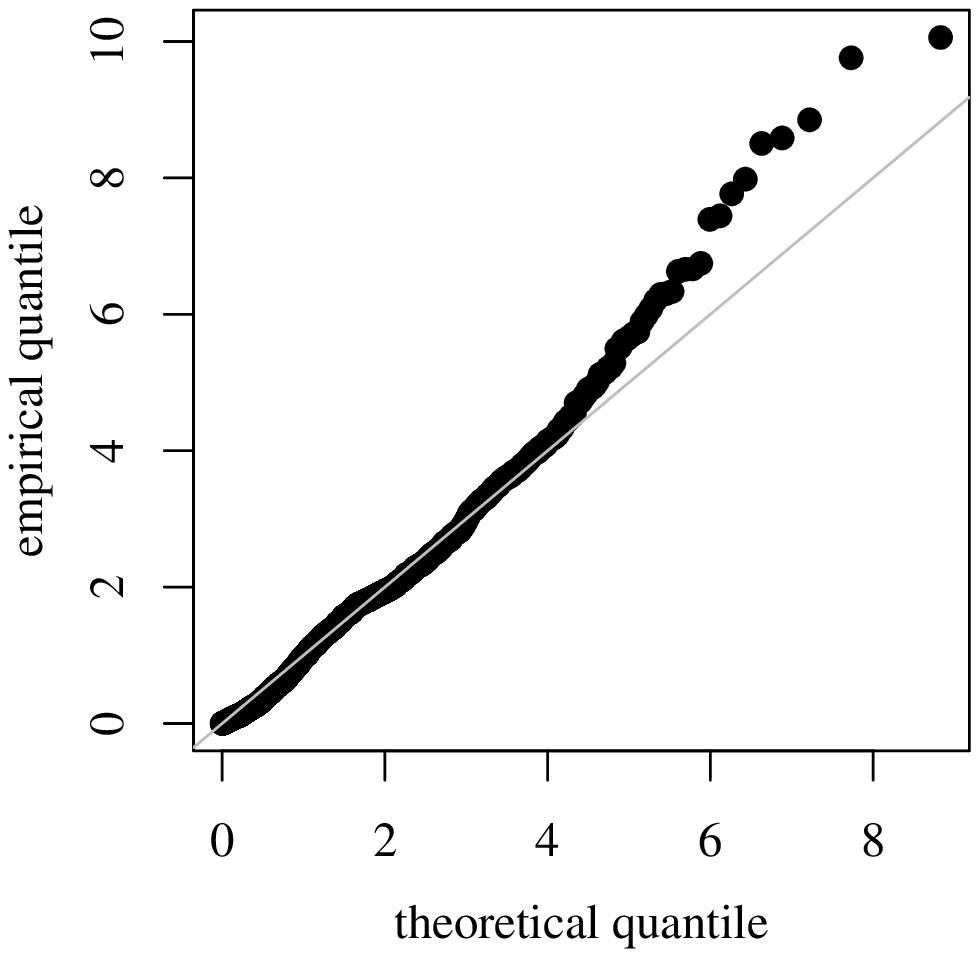}}
 \caption{\small {QQ plot of the $\text{Re}(Y_{1t},Y_{2t})$ residuals for the indicated models.}}
\label{fig:qqplots}
\end{figure}

\begin{figure}[!ht]
\centering
\subfigure[Log-normal]{\includegraphics[height=5.5cm,width=5.5cm]{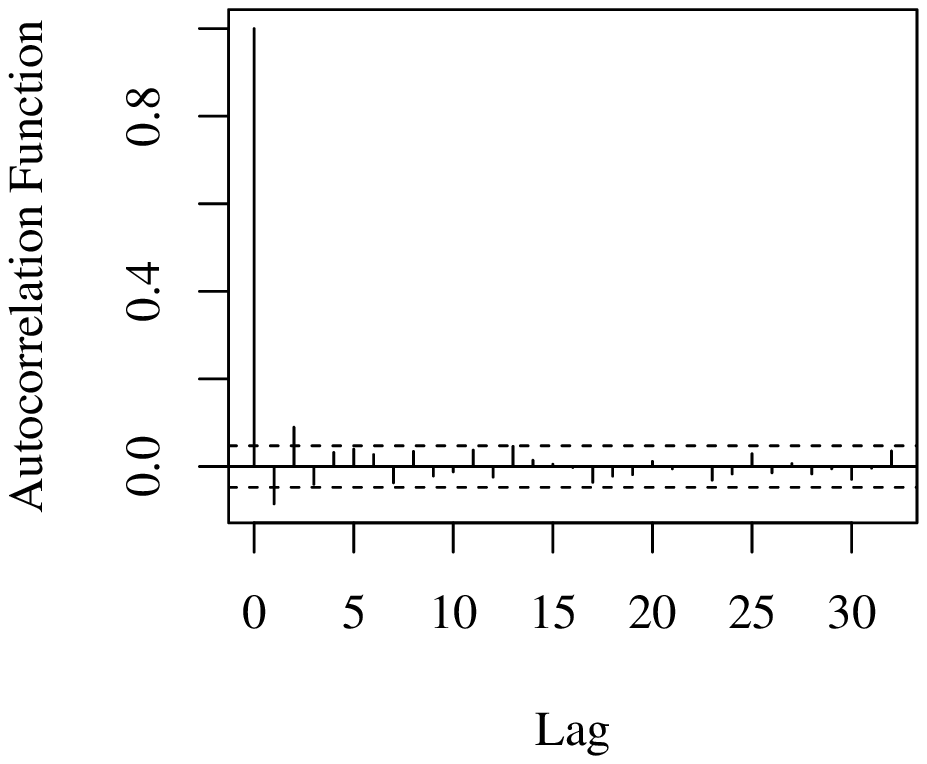}}
\subfigure[Log-normal]{\includegraphics[height=5.5cm,width=5.5cm]{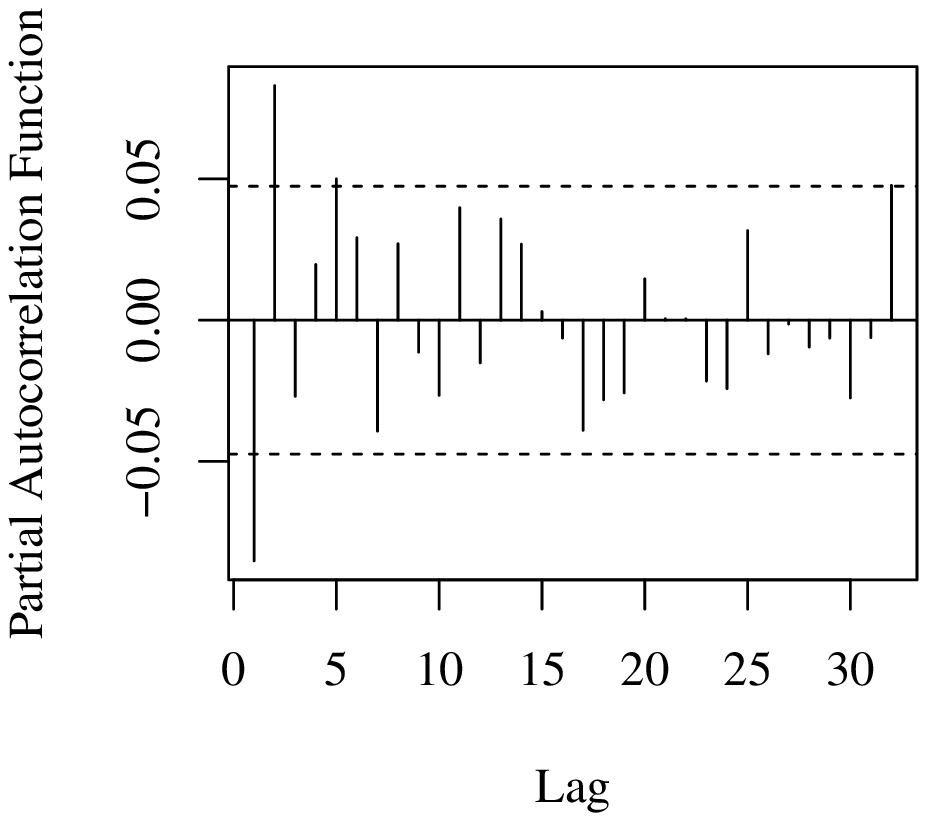}}
\subfigure[Log-hyperbolic]{\includegraphics[height=5.5cm,width=5.5cm]{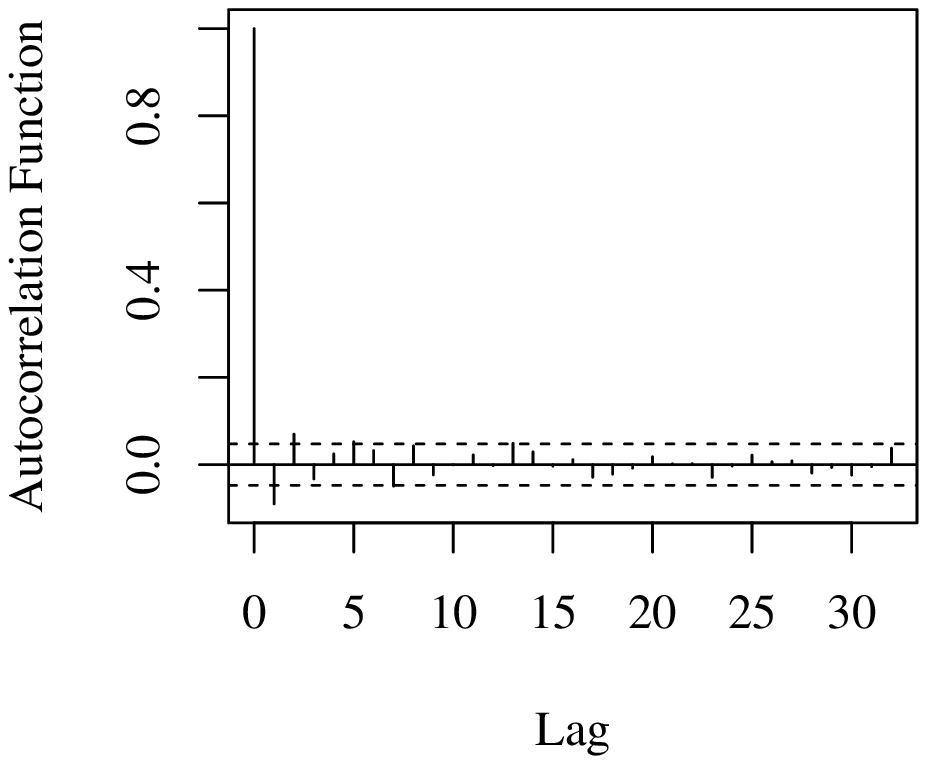}}
\subfigure[Log-hyperbolic]{\includegraphics[height=5.5cm,width=5.5cm]{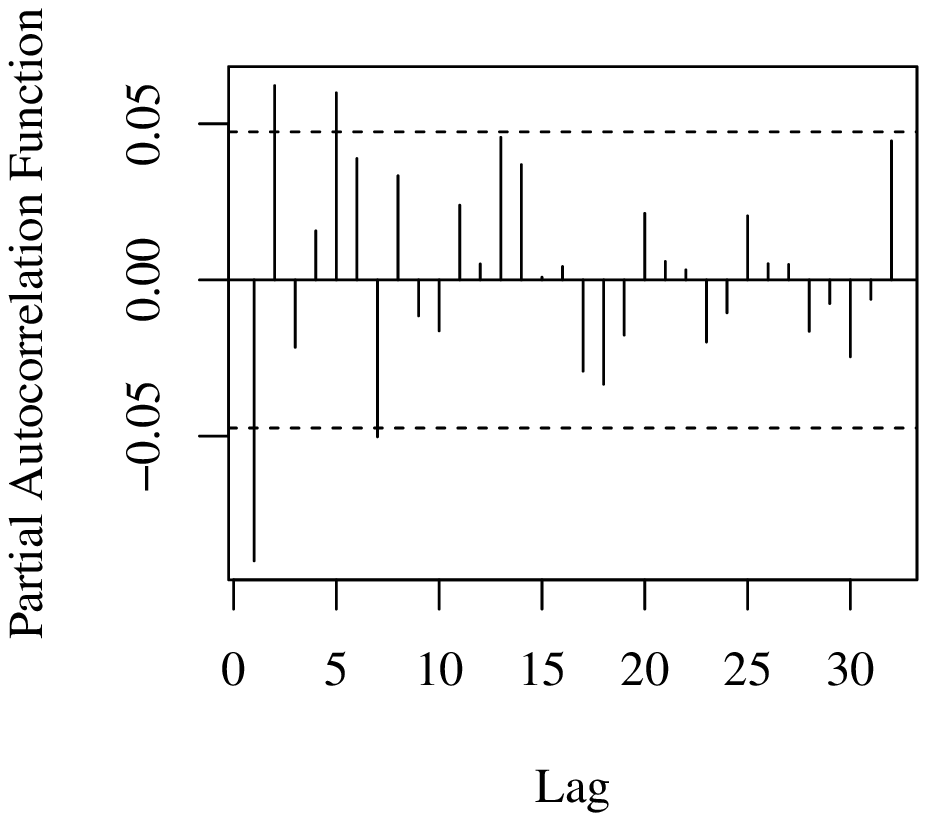}}
\caption{ \small ACF and PACF of the $\text{Re}(Y_{1t},Y_{2t})$ residuals corresponding to the log-normal and log-Laplace models.}
\label{fig:resid_acf}
\end{figure}

\begin{figure}[!ht]
\vspace{-0.25cm}
\centering
{\includegraphics[scale=0.6]{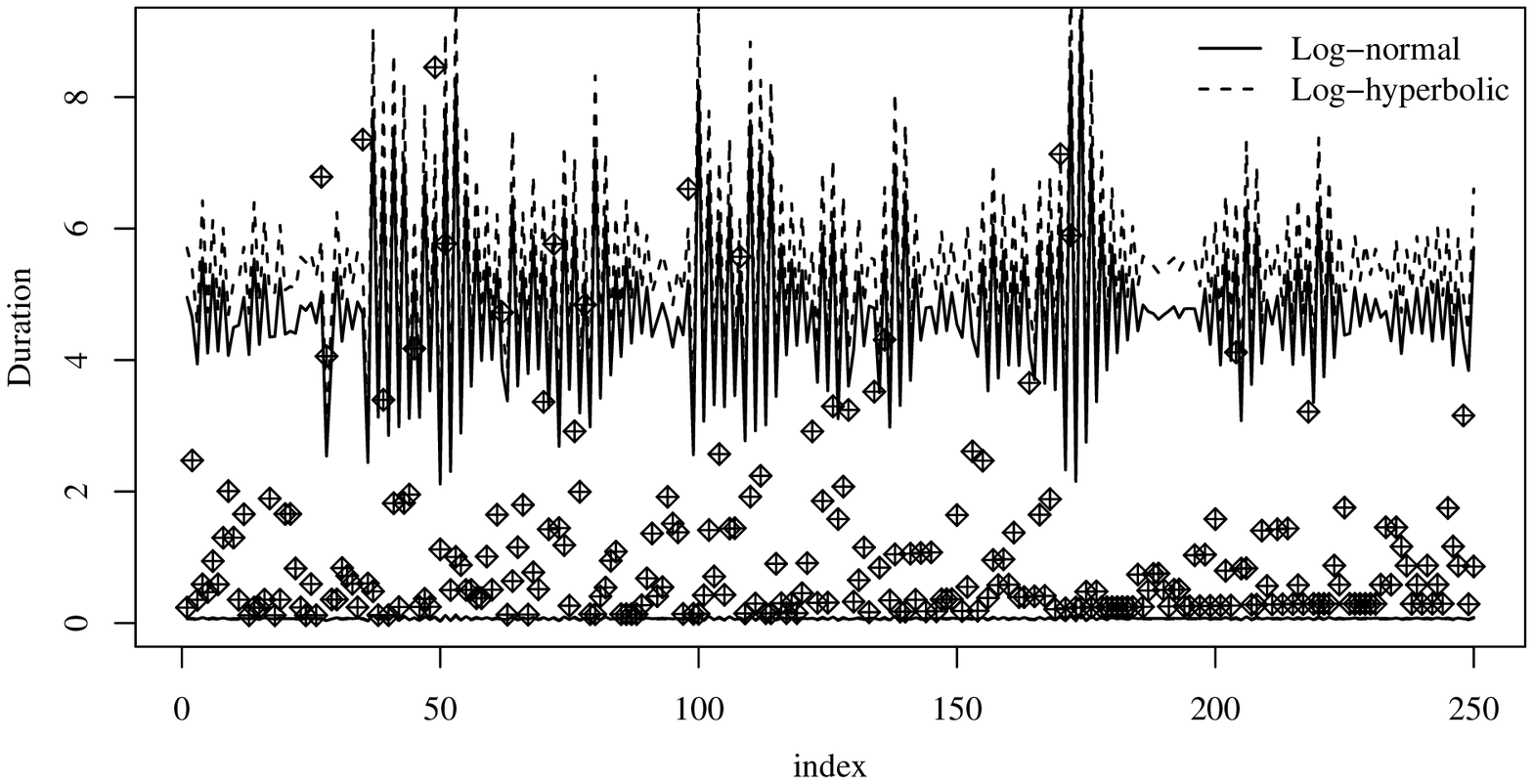}}
{\includegraphics[scale=0.6]{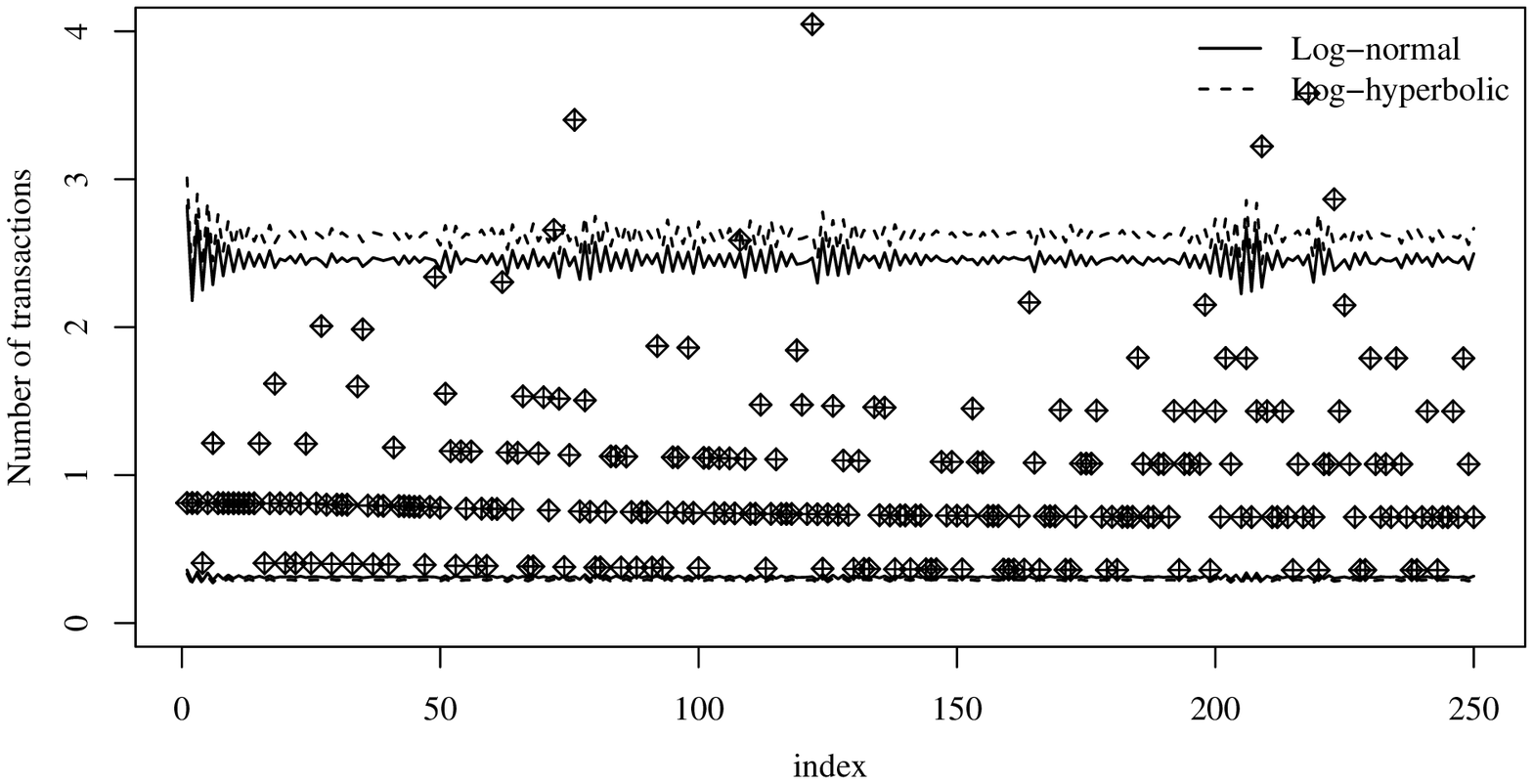}}
\vspace{-0.2cm}
\caption{{95\% prediction intervals from the log-normal and log-Laplace models for the BAC data.}}\label{figfore:predic}
\end{figure}

Figure~\ref{fig:qqplots} presents the QQ plots of the $\text{Re}(Y_{1t},Y_{2t})$ residuals for the models considered in Tables \ref{tab:fit_bac1} and \ref{tab:estima_bac1}. From the QQ plots, we observe that the log-normal and log-Laplace models provide good fits. In Figure~\ref{fig:resid_acf}, we present the autocorrelation function (ACF) and partial autocorrelation function (PACF) of the $\text{Re}(Y_{1t},Y_{2t})$ residuals corresponding to the log-normal and log-Laplace models. The ACF and PACF plots of the $\text{Re}(Y_{1t},Y_{2t})$ residuals indicate that the these models produces non-autocorrelated residuals. Figure \ref{figfore:predic} presents the 95\% prediction intervals for the log-normal and log-Laplace models with the BAC data. We considered two-third of data for in-sample (estimation), which corresponds to 1,139 observations. The remaining one-third of data were considered for out-of-sample (forecasting), which corresponds to 569 observations. Figure \ref{figfore:predic} shows only the first 250 observations. Overall, the results show that (96.66\%, 95.43\%) and (97.72\%, 96.31\%) of the observations of ($Y_{1t}$, $Y_{2t}$) are within the limits of the prediction interval for the log-normal and log-Laplace models, respectively. Therefore, the these models provide values closer to the nominal level of 95\%.

\section{Concluding remarks}\label{sec:05}
In this paper, we have proposed a bivariate autoregressive conditional duration model for dealing with bivariate high frequency financial duration data. In particular, we have developed autoregressive conditional duration models based on bivariate log-symmetric distributions, which are useful for modeling strictly positive, asymmetric and light- and heavy-tailed data such as the transaction-level high-frequency financial data. A Monte Carlo simulation has been carried out for evaluating the performance of the maximum likelihood estimates and the evaluation of a form of residuals. We have applied the proposed models to a real financial duration data set corresponding to price durations and number of market events of Bank of America Corporation stocks. The results demonstrate that the proposed models are practical and useful for joint modeling of duration and number of transactions, providing good results both in terms of model fitting and forecasting ability. As part of future research, it will be of interest to explore semiparametric frameworks. Furthermore, influence diagnostic tools can be developed; see \cite{saulolla:19}. We are currently working in these directions and hope to report the findings in a future paper.

\section*{{Funding}}
{Helton Saulo gratefully acknowledges financial support from CNPq, Brazil.}

\section*{{Conflict of Interest}}
The authors declare no conflict of interest.

\normalsize


\end{document}